\documentclass[aps]{revtex4}
\usepackage{amsfonts}
\usepackage{amsmath}
\usepackage{amssymb,epsf}

\begin{document}

\title{Geometrothermodynamics of black holes \\
in Lovelock gravity with a nonlinear electrodynamics}
\author{S. H. Hendi$^{1,2}$\footnote{email address: hendi@shirazu.ac.ir} and R. Naderi$^3$\footnote{email address: r.naderi@azaruniv.edu}}
\affiliation{$^1$ Physics Department and Biruni Observatory,
College of Sciences, Shiraz
University, Shiraz 71454, Iran \\
$^2$ Research Institute for Astrophysics and Astronomy of Maragha (RIAAM),
P.O. Box 55134-441, Maragha, Iran \\
$^3$ Department of Physics, Azarbaijan Shahid Madani University, Tabriz
53714-161, Iran}

\begin{abstract}
The objective of the present paper is to analyze the phase
transition of asymptotically anti-de Sitter (AdS) black hole
solutions in Lovelock gravity in the presence of nonlinear
electrodynamics. First, we present the asymptotically AdS black
hole solutions for two classes of the Born-Infeld type of
nonlinear electrodynamics coupled with Einstein, Gauss-Bonnet and
third order Lovelock gravity, separately. Then, in order to
discuss the phase transition, we calculate both the heat capacity
and the Ricci scalar of the thermodynamical line element. We
present a comparison between the singular points of the Ricci
scalar using Geometrothermodynamics method and the corresponding
vanishing points of the heat capacity in the canonical ensemble.
In addition, we discuss the effects of both Lovelock and nonlinear
electrodynamics on the phase transition points.
\end{abstract}

\maketitle


\section{Introduction}

One of the interesting subjects for studying a thermodynamical
system is phase transitions. In order to investigate the phase
transitions, one may use the micro canonical, canonical and/or
grand canonical ensembles.

On the other hand, phase transitions of a thermodynamical systems
may be investigated by using of the concept of geometry in
thermodynamics (the so-called Geometrothermodynamics with GTD
abbreviation). In this method one may define a thermodynamical
line element and curvature can be interpreted as a system
interaction. One can study the phase transitions by obtaining the
curvature singularities of the thermodynamical metric. Although
the equivalence of these two methods; i.e., the roots of the heat
capacity in the canonical ensemble (computed by using standard
Black hole thermodynamics) and the curvature singularities of the
thermodynamical metric (calculated in GTD approach) has been
checked for asymptotically AdS black holes \cite{Quevedo6}, it is
not in general on firm ground \cite{Equivalence} and so an
investigation of the validity of such an equivalence may be
worthwhile, especially in cases with nonlinear electrodynamics
(NLED) sources.

At first, Weinhold defined the second derivatives of the internal
energy with respect to entropy and other extensive quantities
(such as the electric charge) of a thermodynamical system
to introduce a Riemannian metric \cite{Weinhold1,Weinhold2}%
\begin{equation}
g^{W}=\frac{\partial ^{2}M}{\partial X^{i}\partial X^{j}}dX^{i}dX^{j},\text{
\ }X^{i}=X^{i}(S,Q).  \label{Weinhold}
\end{equation}
Then, Ruppeiner introduced another metric, in which the Riemannian
metric was defined as the negative Hessian of the entropy with
respect to the internal energy and other extensive quantities of a
thermodynamical system \cite{Ruppeiner1,Ruppeiner2}
\begin{equation}
g^{R}=\frac{\partial ^{2}S}{\partial Y^{i}\partial Y^{j}}dY^{i}dY^{j},\text{
\ }Y^{i}=Y^{i}(M,Q).  \label{Ruppeiner}
\end{equation}
It was shown that both the Ruppeiner and Weinhold metrics are
conformally related to each other, where temperature is the
conformal factor \cite{ERconformal}. Recently, it was shown that
the phase transition points of the heat capacity do not match
those in the Weinhold and Ruppeiner metrics. Quevedo has
established a new Legendre-invariant thermodynamical line element
to overcome this problem
\cite{Quevedo1,Quevedo2,Quevedo3,Quevedo4,Quevedo5,Quevedo6,Quevedo7}.
Although there is more than one Legendre-invariant metric, in this
paper we use the simplest Legendre-invariant generalization of the
Weinhold metric, $g^{W}$, which can be written as
\begin{equation}
g=Mg^{W}=M\frac{\partial ^{2}M}{\partial X^{a}\partial X^{b}}dX^{a}dX^{b},
\label{Wein}
\end{equation}%
where $X^{a}=\left\{ S,Q\right\} $ and its Legendre invariant
metric can be written in terms of the Ruppeiner metric, $g^{R}$,
as
\begin{equation}
g=MTg^{R}=-\frac{M}{\left( \frac{\partial S}{\partial M}\right) }\frac{%
\partial ^{2}S}{\partial Y^{a}\partial Y^{b}}dY^{a}dY^{b},  \label{Rupp}
\end{equation}%
where $Y^{a}=\left\{ M,Q\right\} $.

In this paper, we follow the Quevedo method to study GTD in
Einstein, Gauss-Bonnet (GB) and third order Lovelock (TOL)
gravities \cite{Lovelock}. Besides, we consider the recently
proposed Born-Infeld (BI) type models of NLED as a source
\cite{HendiJHEP} (for motivations of Lovelock and the BI type
models, we refer the reader to
\cite{HendiMahmudi,HendiDehghani,HendiAnn1,HendiAnn2}). We should
note that although one can study the constant curvature spacetimes
with positive, zero or negative values for $\Lambda$,  the case of
a negative cosmological constant is of interest for studies of the
AdS/CFT correspondence\cite{Maldacena}. Besides, the Hawking-Page
phase transition can be interpreted using the AdS/CFT
correspondence. So, hereafter, we are interested only in
asymptotically AdS solutions.

The plan of the paper is as follows. In Sec. \ref{Sol}, we give a
brief discussion of the asymptotically AdS black hole solutions of
Einstein, GB and TOL gravities in the presence of NLED. Section
\ref{Therm} is devoted to the calculation of conserved and
thermodynamic quantities and a check of the first law of
thermodynamics. Then, we discuss the thermal stability of the
solutions by calculating the heat capacity. We use the concept of
geometry in thermodynamics to study the phase transition. We also
compare GTD results with the canonical ensemble stability
criterion. Finally, we end with some conclusions.

\section{Asymptotically AdS Black hole solutions of Lovelock gravity with BI type NLED \label{Sol}}

The Lagrangian of the Lovelock gravity coupled with a NLED source
is written as \cite{Lovelock}
\begin{eqnarray}
\pounds  &=&\mathcal{L}_{Lovelock}+\mathcal{L}(\mathcal{F}),  \label{Lagrangian} \\
\mathcal{L}_{Lovelock} &=&\sum_{i=0}\alpha _{i}\mathcal{L}_{i},
\end{eqnarray}%
where $\alpha _{i}$ and $\mathcal{L}_{i}$ indicate the
coefficients and Lagrangians of Lovelock gravity, respectively,
and $\mathcal{L}(\mathcal{F})$ is the Lagrangian of NLED. In this
paper, we regard the Lovelock gravity up to the fourth term. More
explicitly, $\mathcal{L}_{0}=-2\Lambda $, in which $\Lambda $ is
the negative cosmological constant, $\mathcal{L}_{1}=R$ denotes
the Ricci scalar, $ \mathcal{L}_{2}= R_{\mu \nu \gamma \delta
}R^{\mu \nu \gamma \delta }-4R_{\mu \nu }R^{\mu \nu }+R^{2}$ is
the Lagrangian of Gauss-Bonnet gravity and the Lagrangian of third
order Lovelock gravity, $\mathcal{L}_{3}$, is
\begin{eqnarray}
\mathcal{L}_{3} &=&2R^{\mu \nu \sigma \kappa }R_{\sigma \kappa \rho \tau }R_{%
\phantom{\rho \tau }{\mu \nu }}^{\rho \tau }+8R_{\phantom{\mu \nu}{\sigma
\rho}}^{\mu \nu }R_{\phantom {\sigma \kappa} {\nu \tau}}^{\sigma \kappa }R_{%
\phantom{\rho \tau}{ \mu \kappa}}^{\rho \tau }+  \notag \\
&&24R^{\mu \nu \sigma \kappa }R_{\sigma \kappa \nu \rho }R_{%
\phantom{\rho}{\mu}}^{\rho }+3RR^{\mu \nu \sigma \kappa }R_{\sigma \kappa
\mu \nu }+24R^{\mu \nu \sigma \kappa }R_{\sigma \mu }R_{\kappa \nu }+  \notag
\\
&&16R^{\mu \nu }R_{\nu \sigma }R_{\phantom{\sigma}{\mu}}^{\sigma }-12RR^{\mu
\nu }R_{\mu \nu }+R^{3}.  \label{L3}
\end{eqnarray}

In order to consider an appropriate model of NLED, we take into
account the recently proposed BI type models of NLED which were
introduced by Hendi (exponential form) \cite{HendiJHEP} and Soleng
(logarithmic form) \cite{Soleng} with the following explicit forms
\begin{equation}
\mathcal{L}(\mathcal{F})=\left\{
\begin{array}{ll}
\beta ^{2}\left( \exp (-\frac{\mathcal{F}}{\beta ^{2}})-1\right) ,
& \text{ENEF} \\
-8\beta ^{2}\ln \left( 1+\frac{\mathcal{F}}{8\beta ^{2}}\right) ,
& \text{LNEF}
\end{array}
\right. ,  \label{LNon}
\end{equation}
where $\beta$ denotes the nonlinearity parameter and the Maxwell
invariant is $\mathcal{F}=F_{ab}F^{ab}$, in which $F_{ab}=\partial
_{a}A_{b}-\partial _{b}A_{a}$ is the electromagnetic field tensor
and $A_{a}$ is the gauge potential. We should note that for large
values of $\beta$, the Lagrangians of Eq. (\ref{LNon}) reduce to
the linear Maxwell Lagrangian.

Taking into account both gravitational ($g^{a b}$) and
electromagnetic ($A^{a}$) fields, and using the Euler-Lagrange
equation, the field equations of Lovelock gravity in the presence
of NLED are described by \cite{Muller}
\begin{equation}
\alpha _{0}G_{\mu \nu }^{(0)}+\alpha _{1}G_{\mu \nu }^{(1)}+\alpha
_{2}G_{\mu \nu }^{(2)}+\alpha _{3}G_{\mu \nu
}^{(3)}=\frac{1}{2}g_{\mu \nu }\mathcal{L}(\mathcal{F})-2F_{\mu
\lambda }F_{\nu }^{\;\lambda }\mathcal{L}_{\mathcal{F}},
\label{FE1}
\end{equation}%
and
\begin{equation}
\partial _{\mu }\left( \sqrt{-g}\mathcal{L}_{\mathcal{F}}F^{\mu \nu }\right) =0,
\label{FE2}
\end{equation}%
where $\alpha _{0}=\alpha _{1}=1$, the arbitrary coefficients
$\alpha _{2}$ and $\alpha _{3}$ are related to GB and TOL gravity
and
\begin{eqnarray}
&&G_{\mu \nu }^{(0)}=-\frac{1}{2}g_{\mu \nu }\mathcal{L}_{0},  \label{Love1}
\\
&&G_{\mu \nu }^{(1)}=R_{\mu \nu }-\frac{1}{2}g_{\mu \nu }\mathcal{L}_{1}, \\
&&G_{\mu \nu }^{(2)}=2(R_{\mu \sigma \kappa \tau }R_{\nu }^{\phantom{\nu}%
\sigma \kappa \tau }-2R_{\mu \rho \nu \sigma }R^{\rho \sigma }-2R_{\mu
\sigma }R_{\phantom{\sigma}\nu }^{\sigma }+RR_{\mu \nu })-\frac{1}{2}g_{\mu
\nu }\mathcal{L}_{2},  \label{Love22}
\end{eqnarray}%
\begin{eqnarray}
G_{\mu \nu }^{(3)} &=&-3(4R^{\tau \rho \sigma \kappa }R_{\sigma \kappa
\lambda \rho }R_{\phantom{\lambda }{\nu \tau \mu}}^{\lambda }-8R_{%
\phantom{\tau \rho}{\lambda \sigma}}^{\tau \rho }R_{\phantom{\sigma
\kappa}{\tau \mu}}^{\sigma \kappa }R_{\phantom{\lambda }{\nu \rho \kappa}%
}^{\lambda }+  \notag \\
&&2R_{\nu }^{\phantom{\nu}{\tau \sigma \kappa}}R_{\sigma \kappa \lambda \rho
}R_{\phantom{\lambda \rho}{\tau \mu}}^{\lambda \rho }-R^{\tau \rho \sigma
\kappa }R_{\sigma \kappa \tau \rho }R_{\nu \mu }+  \notag \\
&&8R_{\phantom{\tau}{\nu \sigma \rho}}^{\tau }R_{\phantom{\sigma
\kappa}{\tau \mu}}^{\sigma \kappa }R_{\phantom{\rho}\kappa }^{\rho }+8R_{%
\phantom {\sigma}{\nu \tau \kappa}}^{\sigma }R_{\phantom {\tau \rho}{\sigma
\mu}}^{\tau \rho }R_{\phantom{\kappa}{\rho}}^{\kappa }+  \notag \\
&&+4R_{\nu }^{\phantom{\nu}{\tau \sigma \kappa}}R_{\sigma \kappa \mu \rho
}R_{\phantom{\rho}{\tau}}^{\rho }-4R_{\nu }^{\phantom{\nu}{\tau \sigma
\kappa }}R_{\sigma \kappa \tau \rho }R_{\phantom{\rho}{\mu}}^{\rho }+  \notag
\\
&&4R^{\tau \rho \sigma \kappa }R_{\sigma \kappa \tau \mu }R_{\nu \rho
}+2RR_{\nu }^{\phantom{\nu}{\kappa \tau \rho}}R_{\tau \rho \kappa \mu }+8R_{%
\phantom{\tau}{\nu \mu \rho }}^{\tau }R_{\phantom{\rho}{\sigma}}^{\rho }R_{%
\phantom{\sigma}{\tau}}^{\sigma }  \notag \\
&&-8R_{\phantom{\sigma}{\nu \tau \rho }}^{\sigma }R_{\phantom{\tau}{\sigma}%
}^{\tau }R_{\mu }^{\rho }-8R_{\phantom{\tau }{\sigma \mu}}^{\tau \rho }R_{%
\phantom{\sigma}{\tau }}^{\sigma }R_{\nu \rho }-  \notag \\
&&4RR_{\phantom{\tau}{\nu \mu \rho }}^{\tau }R_{\phantom{\rho}\tau }^{\rho
}+4R^{\tau \rho }R_{\rho \tau }R_{\nu \mu }-8R_{\phantom{\tau}{\nu}}^{\tau
}R_{\tau \rho }R_{\phantom{\rho}{\mu}}^{\rho }+  \notag \\
&&4RR_{\nu \rho }R_{\phantom{\rho}{\mu }}^{\rho }-R^{2}R_{\nu \mu })-\frac{1%
}{2}\mathcal{L}_{3}g_{\mu \nu },  \label{Love3}
\end{eqnarray}%
and $\mathcal{L}_{\mathcal{F}}=\frac{d
\mathcal{L}(\mathcal{F})}{d\mathcal{F}}$.

Now, we should consider a suitable metric and study the effects of
both the Lovelock and NLED terms. The ($n+1$)-dimensional line
element of a spherically symmetric spacetime may be written as
\begin{equation}
ds^{2}=-N(r)f(r)dt^{2}+\frac{dr^{2}}{f(r)}+r^{2}\left( d\theta
_{1}^{2}+\sum\limits_{i=2}^{n-1}\prod\limits_{j=1}^{i-1}\sin ^{2}\theta
_{j}d\theta _{i}^{2}\right) ,  \label{Metric}
\end{equation}
and hereafter we suppose that the volume of a ($n-1$)-dimensional
$t=constant$ and $r=constant$ hypersurface is $V_{n-1}$. The
nonzero components of the electromagnetic field tensor in
arbitrary ($n+1$)-dimensions may be written as
\cite{HendiAnn1,HendiAnn2}
\begin{equation}
F_{tr}=-F_{rt}=\frac{q}{r^{n-1}}\times \left\{
\begin{array}{cc}
\exp \left( -\frac{L_{W}}{2}\right) , & ENEF \\
\frac{2}{1+\Gamma }, & LNEF
\end{array}
\right. ,  \label{Ftr}
\end{equation}
where
\begin{eqnarray*}
L_{W} &=&LambertW\left( \frac{4q^{2}}{\beta ^{2}r^{2n-2}}\right) , \\
\Gamma  &=&\sqrt{1+\frac{q^{2}}{\beta ^{2}r^{2n-2}}},
\end{eqnarray*}%
and $q$ is an integration constant related to the electric charge.
The metric function of Einstein gravity that satisfies the
gravitational field equation ($\alpha _{2}=\alpha _{3}=0$) is
\cite{HendiAnn1,HendiAnn2}
\begin{equation}
f_{E}(r)=1-\frac{2\Lambda r^{2}}{n(n-1)}-\frac{m}{r^{n-2}}+\Sigma
, \label{fE}
\end{equation}
where
\begin{equation}
\Sigma =\left\{
\begin{array}{cc}
-\frac{\beta ^{2}r^{2}}{n(n-1)}+\frac{2q\beta }{(n-1)r^{n-2}}\int \frac{%
1-L_{W}}{\sqrt{L_{W}}}dr, & \text{ENEF}\vspace{0.1cm} \\
-\frac{16\beta ^{2}r^{2}}{n(n-1)}-\frac{8\beta ^{2}\ln (2)}{n(n-1)}+\frac{8}{%
(n-1)r^{n-2}}\int \frac{\frac{q^{2}}{\Gamma -1}-\beta ^{2}\ln \left( \frac{%
\beta ^{2}r^{2n-2}\left( \Gamma -1\right) }{q^{2}}\right)
}{r^{n-1}}dr, & \text{LNEF}
\end{array}
\right. ,  \label{Sig}
\end{equation}
and $m$ is an integration constant related to the finite mass.
Looking at last the term of Eq. (\ref{Sig}), one may think that,
for some specific limits, this term behaves like a mass term
(proportional to $r^{2-n}$); however, this is not true. We should
note that the only integration constant of the field equation was
labeled with $m$ in Eq. (\ref{fE}) and we cannot adjust $q$ and
$\beta$ to obtain a constant value for the integration part of the
last term in Eq. (\ref{Sig}) for arbitrary $r$
\cite{HendiAnn1,HendiAnn2}.

The metric function for GB gravity ($\alpha _{2}\neq 0$, $\alpha
_{3}=0$) can be written as \cite{HendiMahmudi}
\begin{equation}
f_{GB}(r)=1+\frac{r^{2}}{2\alpha }\left( 1-\sqrt{\Psi _{GB}}\right) ,
\label{fGB}
\end{equation}
where
\begin{eqnarray}
\Psi _{GB} &=&1+\frac{8\alpha \Lambda }{n(n-1)}+\frac{4\alpha
m}{r^{n}}+\frac{4\alpha \beta ^{2}\Upsilon _{n}}{n\left(
n-1\right) },  \label{Psi} \\ \\ \Upsilon _{n} &=&\left\{
\begin{array}{cc}
1+\frac{2nq}{\beta r^{n}} \int
\frac{L_{W}-1}{\sqrt{L_{W}}}dr,\;&\;\text{ENEF} \\
\\ \frac{8(n-1)}{n}\left[ \frac{\left( 2n-1\right) (\Gamma
-1)}{n-1}-\frac{n\ln \left( \frac{1+\Gamma }{2}\right)
}{n-1}+\frac{\left( n-1\right) \left( 1-\Gamma ^{2}\right)
\mathcal{H}}{n-2}\right] ,\;&\;\text{LNEF}
\end{array}
\right. ,  \label{Gam} \\
\mathcal{H} &=& {}_{2}{F_{1}}  \left(
\frac{1}{2},\frac{n-2}{2n-2};  \frac{3n-4}{2n-2};\left( 1-\Gamma
^{2}\right) \right), \label{H}
\end{eqnarray}%
in which ${}_{p}{F_{q}}(a,b;c;d)$ is the hypergeometric function.
Since the inclusion of the mentioned NLED models in the Lagrangian
of third order Lovelock gravity with two free parameters
$\alpha_{2}$ and $\alpha_{3}$ makes calculations considerably more
complicated, we use a subclass of the Lovelock gravity with a
relation between the various Lovelock coefficients. In other
words, in our model, one can choose a special case in which
$\alpha_{3}$ is a function of $\alpha_{2}$ to simplify the
complicated TOL field equation and its solutions. Thus, the metric
function for TOL gravity may be obtained as \cite{HendiDehghani}
\begin{equation}
f_{TOL}(r)=1+\frac{r^{2}}{\alpha }\left[ 1-\Psi _{TOL}^{1/3}\right] ,
\label{fTOL}
\end{equation}
where
\begin{equation}
\Psi _{TOL}=1+\frac{{3\alpha m}}{{{r^{n}}}}+\frac{{6\alpha \Lambda }}{{n(n-1)%
}}+\frac{3{\alpha {\beta ^{2}}}\Upsilon _{n}}{n\left( n-1\right) },
\label{Psi2}
\end{equation}
and $N(r)=C$ for all the mentioned branches of the Lovelock
gravity. Hereafter, we choose $N(r)=C=1$ without loss of
generality. In the above equations, we have set $\alpha
_{2}=\frac{\alpha }{(n-2)(n-3)}$ and $\alpha _{3}=\frac{\alpha
^{2}}{3(n-2)(n-3)(n-4)(n-5)}$ for further simplification. It has
been shown that these solutions may be interpreted as black hole
solutions with various horizon properties depending on the values
of the nonlinearity parameter $\beta $ (see
\cite{HendiJHEP,HendiAnn1} for more details). Using the series
expansion for large distances ($r>>1$), one can show that these
solutions are asymptotically AdS (with an effective cosmological
constant). Besides, we should note that in general, there are no
restrictions on the Lovelock coefficient $\alpha$. Although there
are a limited number of published papers covering negative values
for the Lovelock coefficient \cite{NegAlpha}, in this paper, we
restrict ourselves to positive $\alpha$. In addition, in order to
obtain physical solutions with real asymptotical behavior, we
should regard $\alpha \leq -n(n-1)/(8\Lambda)$ for negative
$\Lambda$. Since the geometric properties of the solutions were
discussed before \cite
{HendiJHEP,HendiAnn2,HendiMahmudi,HendiDehghani}, in this paper,
we focus on the thermodynamic stability conditions using GTD
method.

\section{Thermodynamic stability \label{Therm}}

\subsection{Thermodynamic properties}

Taking into account third order Lovelock gravity, it has been
shown that the entropy of the asymptotically flat solutions can be
written as the Wald formula \cite{Wald}
\begin{equation}
S=\frac{1}{4}\int d^{n-1}x\sqrt{\tilde{g}}\left( 1+2\alpha _{2}\tilde{R}%
+3\alpha _{3}\mathcal{\tilde{L}}_{2}\right)  \label{Entb}
\end{equation}%
where $\tilde{g}$ is the determinant of the induced metric
$\tilde{g}_{\mu \nu }$, $\tilde{R}$ and $\mathcal{\tilde{L}}_{2}$
are, respectively, the Ricci scalar and the Lagrangian of GB
gravity for the metric $\tilde{g}_{ab}$ on the $(n-1)$-dimensional
spacelike hypersurface. Calculations show that for our black hole
solutions the entropy may be simplified as
\cite{HendiAnn2,HendiMahmudi,HendiDehghani}
\begin{equation}
S=\frac{V_{n-1}r_{+}^{n-1}}{4}\left\{
\begin{array}{cc}
1, & \text{Einstein} \\
1+{\frac{{2\left( {n-1}\right) }\alpha }{(n-3){r_{+}^{2}}},} & \text{GB} \\
{1+\frac{{2\left( {n-1}\right) }}{{\left( {n-3}\right) }}\frac{\alpha }{{%
r_{+}^{2}}}+\frac{{\left( {n-1}\right) }}{{\left( {n-5}\right) }}\frac{{{%
\alpha ^{2}}}}{{r_{+}^{4}}}}, & \text{TOL}%
\end{array}%
\right. ,  \label{Entropy}
\end{equation}
where for $\alpha \longrightarrow 0$ the area law is recovered, as
is expected. In Eq. (\ref{Entropy}), $r_{+}$ is the event horizon
radius of the black hole solutions. It is notable that one can
obtain Eq. (\ref{Entropy}) from the Gibbs--Duhem relation
\cite{GD}.

Now, we calculate the flux of the electric field at infinity to
obtain the electric charge of the black holes. For the mentioned
BI type NLED theories, we find
\cite{HendiAnn2,HendiMahmudi,HendiDehghani}
\begin{equation}
Q=\frac{V_{n-1}}{4\pi }q,  \label{Charge}
\end{equation}%
which is the same as that in the linear Maxwell theory. Regarding
the temporal Killing null generator $\chi =\partial /\partial t$,
the electric potential $U$, measured at infinity with respect to
the event horizon, is \cite{Gub1,Gub2}
\begin{eqnarray}
U &=&A_{\mu }\chi ^{\mu }\left\vert _{r\rightarrow \infty }-A_{\mu }\chi
^{\mu }\right\vert _{r=r_{+}}  \nonumber \\
&=&\left\{
\begin{array}{ll}
\frac{{\beta {r_{+}}\sqrt{L_{W_{+}}} }}{{2(n-2)(3n-4)}}\left[ (n-1)L_{W_{+}} \zeta _{+}+3n-4 \right] , & \text{ENEF}\vspace{0.1cm} \\
-\frac{{2{\beta ^{2}}r_{+}^{n}}}{{nq}}\left( {{\eta
_{+}}-1}\right) , & \text{LNEF}
\end{array}
\right. ,   \label{U}
\end{eqnarray}
where%
\begin{equation}
\zeta _{+}={}_{1}{F_{1}}\left( 1; {\frac{{5n-6}}{{2(n-1)}}}
;\frac{L_{W_{+}}}{{2(n-1)}}\right) ,  \label{Fzeta}
\end{equation}
\begin{equation}
\eta _{+}={}_{2}{F_{1}}\left( -\frac{1}{2},\frac{{-n}}{{2(n-1) }}
; {\frac{{n-2}}{{2(n-1)}}} ;\left( 1-\Gamma _{+}^{2}\right)
\right) ,  \label{Feta}
\end{equation}
and
\begin{eqnarray*}
L_{W_{+}} &=&LambertW\left( \frac{4q^{2}}{\beta ^{2}r_{+}^{2n-2}}\right) , \\
\Gamma_{+}  &=&\sqrt{1+\frac{q^{2}}{\beta ^{2}r_{+}^{2n-2}}}.
\end{eqnarray*}%

Besides, the Hawking temperature of the black hole may be obtained
by the use of the surface gravity interpretation, yielding
\begin{eqnarray}
T_{E} &=&\frac{-2\Lambda r_{+}^{2}+(n-1)(n-2) -\frac{\Phi }{r_{+}}}{4\pi (n-1)r_{+}},  \label{TempE} \\
T_{GB} &=&\frac{-2\Lambda r_{+}^{4}+(n-1)(n-2) r_{+}^{2}+
(n-1)(n-4) \alpha -\Phi r_{+}}{4\pi
r_{+} (n-1) \left( r_{+}^{2}+2\alpha \right) },  \label{TempGB} \\
T_{TOL} &=&\frac{-6\Lambda r_{+}^{6}+3 (n-1)(n-2)
r_{+}^{4}+3(n-1)(n-4) \alpha r_{+}^{2}+ (n-1)(n-6) \alpha
^{2}-3\Phi r_{+}^{3}}{12\pi r_{+}(n-1) \left( r_{+}^{2}+\alpha
\right) ^{2}}, \label{TempTOL}
\end{eqnarray}
where
\begin{equation*}
\Phi =\left\{
\begin{array}{c}
\beta ^{2}r_{+}^{3}\left( \left[ 1+\left( \frac{2E}{\beta }\right)
^{2}\right] e^{\frac{-2E^{2}}{\beta ^{2}}}-1\right) ,\;\;\text{ENEF} \\
\;8r_{+}^{3}\beta ^{2}\ln \left[ 1-\left( \frac{E}{2\beta }\right)
^{2}\right] +\frac{4r_{+}^{3}E^{2}}{1-\left( \frac{E}{2\beta
}\right) ^{2}},\;\;\text{LNEF}
\end{array}
\right. ,
\end{equation*}
and
\begin{equation}
E=\frac{q}{r_{+}^{n-1}}\times \left\{
\begin{array}{cc}
\exp \left( -\frac{L_{W_{+}}}{2}\right) , & ENEF \\
\frac{2}{1+\Gamma_{+} }, & LNEF
\end{array}
\right. .  \label{E}
\end{equation}

Following the method of \cite{HendiJHEP}, one can find that there
is a critical nonlinearity parameter, $\beta _{c}$, in which the
Hawking temperature is positive definite for $\beta <\beta _{c}$.
For $\beta >\beta _{c}$, there is a minimum value for the horizon
radius of physical black holes, $r_{0}$, in which $T$ is positive
for $r_{+}>r_{0}$.

Regarding the Arnowitt-Deser-Misner (ADM) approach, we find that
the finite mass of a black hole is \cite{Brewin}
\begin{equation}
M=\frac{V_{n-1}}{16\pi }m\left( n-1\right) ,  \label{Mass}
\end{equation}
where $m$ can be calculated from $\left.f(r) \right\vert
_{r=r_{+}}=0$ and, therefore, in general the finite mass depends
on both the Lovelock coefficients and the nonlinearity parameter.

It has been shown that the obtained conserved and thermodynamic
quantities satisfy the first law of thermodynamics \cite
{HendiAnn2,HendiMahmudi,HendiDehghani}
\begin{equation}
dM=TdS+UdQ.  \label{1stLaw}
\end{equation}
In other words, by combining Eq. (\ref{1stLaw}) with Eqs.
(\ref{fE}) [or other metric functions of GB and TOL branches:
(\ref{fGB}) and (\ref{fTOL})], (\ref{Entropy}), (\ref{Charge}) and
(\ref{Mass}), we find that $\left(\frac{\partial M}{\partial
Q}\right)_{S}$ and $\left(\frac{\partial M}{\partial
S}\right)_{Q}$ are the same as those calculated in Eqs. (\ref{U})
and (\ref{TempE}) [ or other temperatures of GB and TOL branches:
(\ref{TempGB}) and (\ref{TempTOL})], respectively.

\subsection{Thermal stability and Geometrothermodynamics}

Now, we investigate the thermal stability in the canonical
ensemble by calculating the heat capacity of the black hole
solutions
\begin{equation}
C_{Q}\equiv T\left( \frac{\partial S}{\partial T}\right) _{Q}=T\left( \frac{%
\partial ^{2}M}{\partial S^{2}}\right) _{Q}^{-1}.  \label{Heat}
\end{equation}%
The root of the heat capacity corresponds to the phase transition,
and the positivity of $C_{Q}$ is a sufficient condition for the
system to be locally stable \cite{Chamblin,Davies}.

\begin{figure}[tbp]
$%
\begin{array}{cc}
\epsfxsize=7.5cm \epsffile{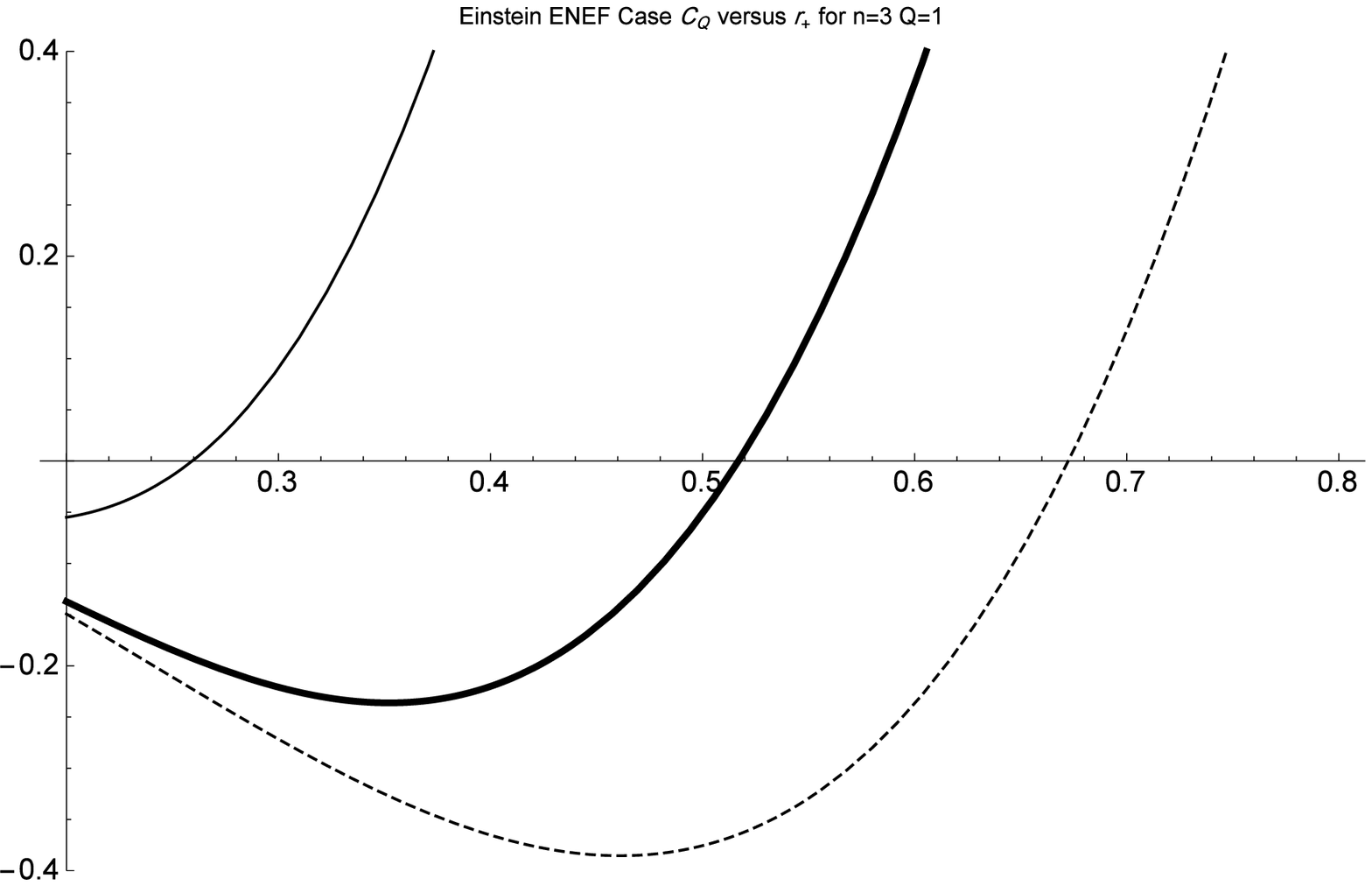} & \epsfxsize=7.5cm
\epsffile{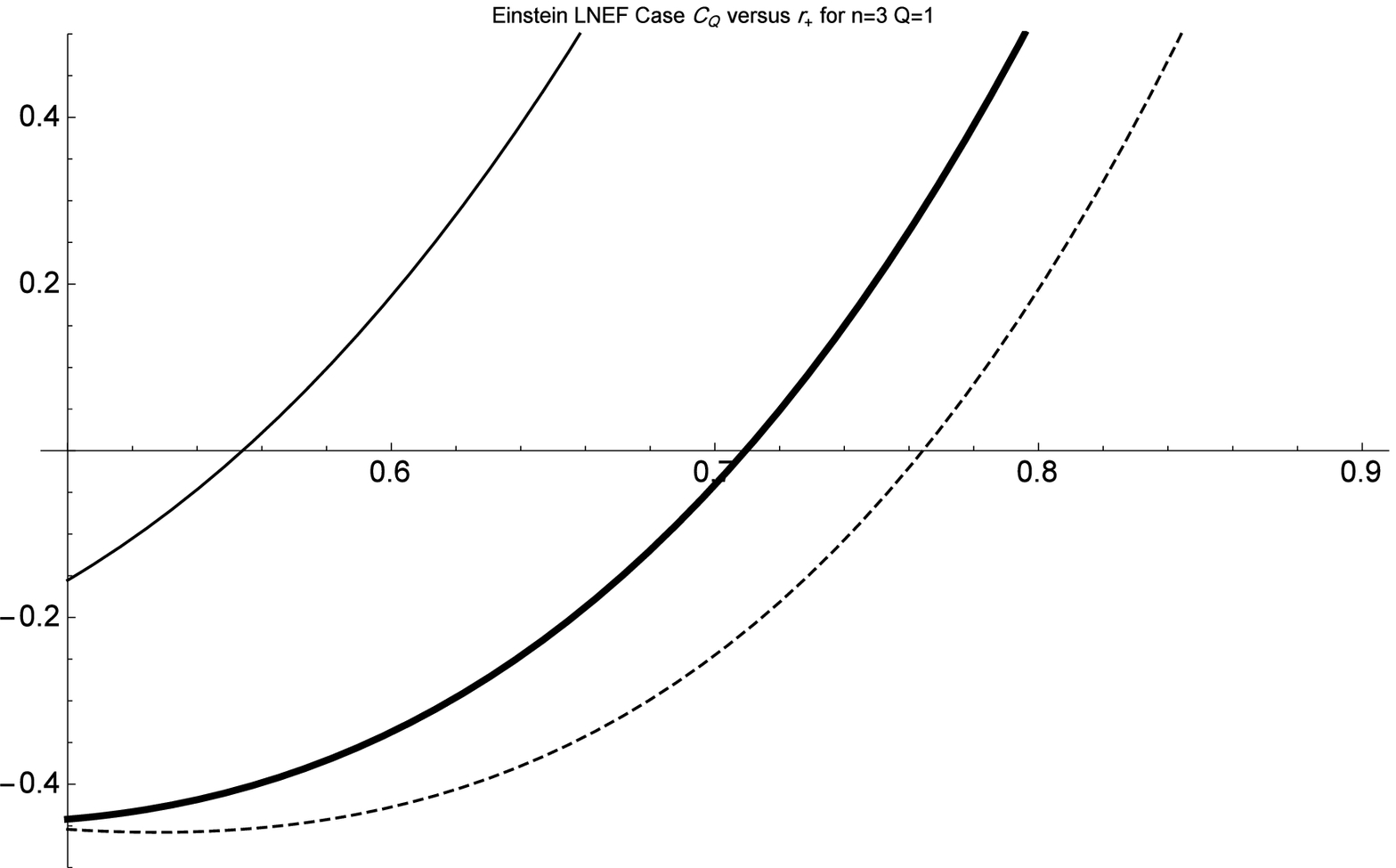} \\
\epsfxsize=7.5cm \epsffile{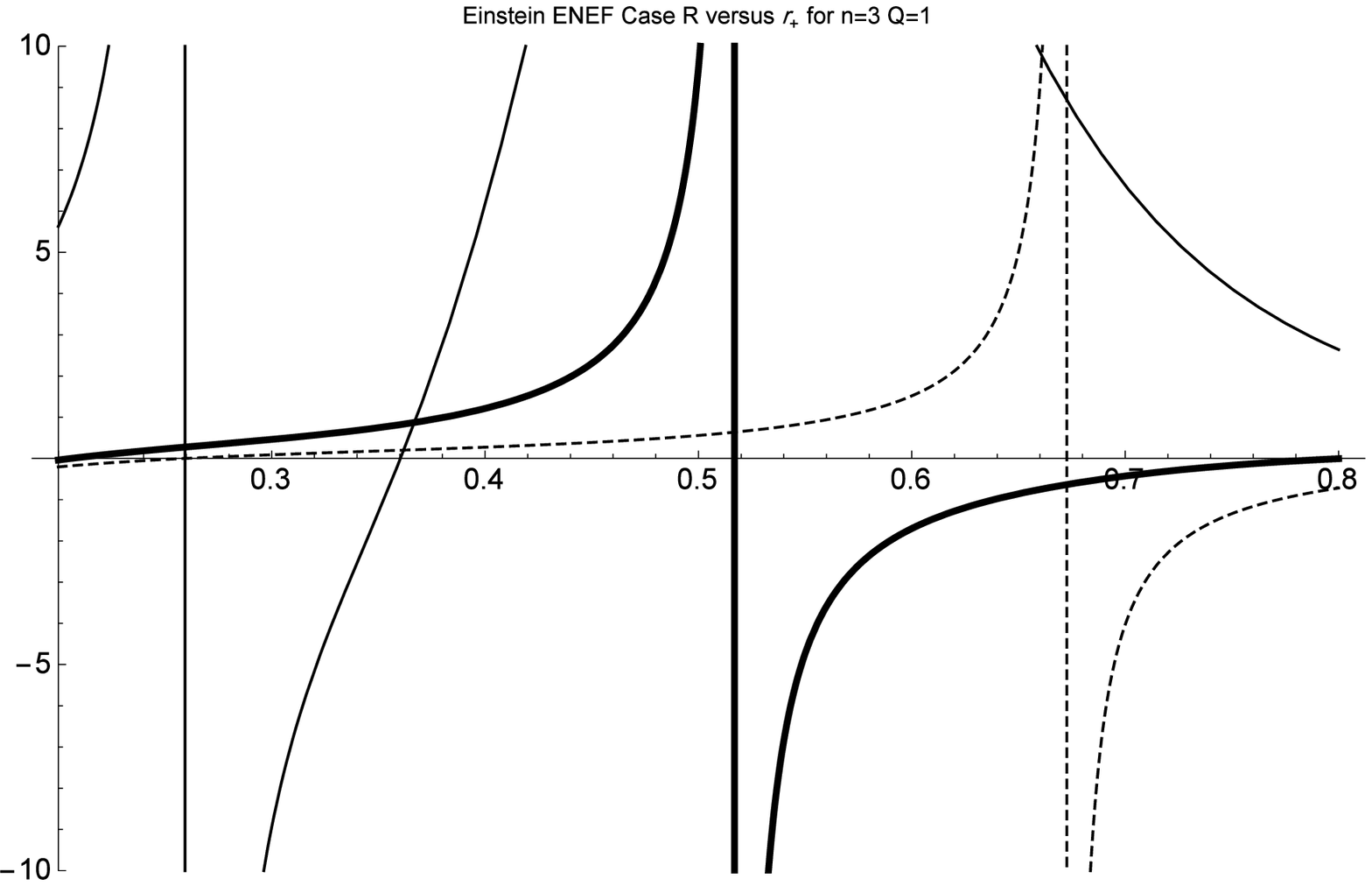} & \epsfxsize=7.5cm
\epsffile{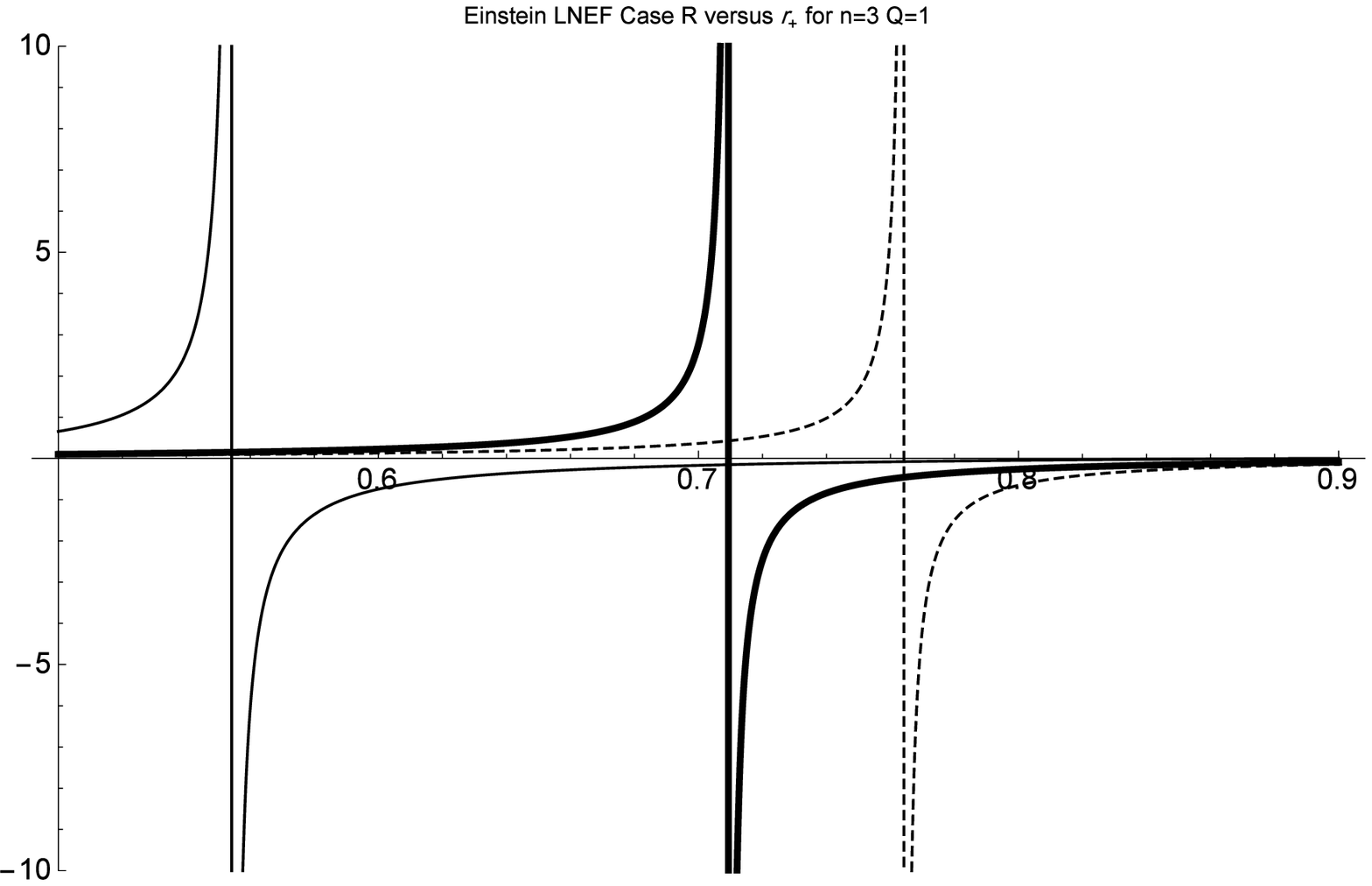}
\end{array}
$%
\caption{\textbf{" Einstein case: ENEF (left) and LNEF (right)
branches:"} Heat capacity (up) and Geometric Ricci scalar (down)
versus $r_{+}$ for $n=3$, $\Lambda=-1$, $Q=1$ and $\protect\beta
=0.5$ (solid line), $\protect\beta =1$ (bold line) and
$\protect\beta =2$ (dashed line). } \label{HEAT1}
\end{figure}

\begin{figure}[tbp]
$%
\begin{array}{cc}
\epsfxsize=7.5cm \epsffile{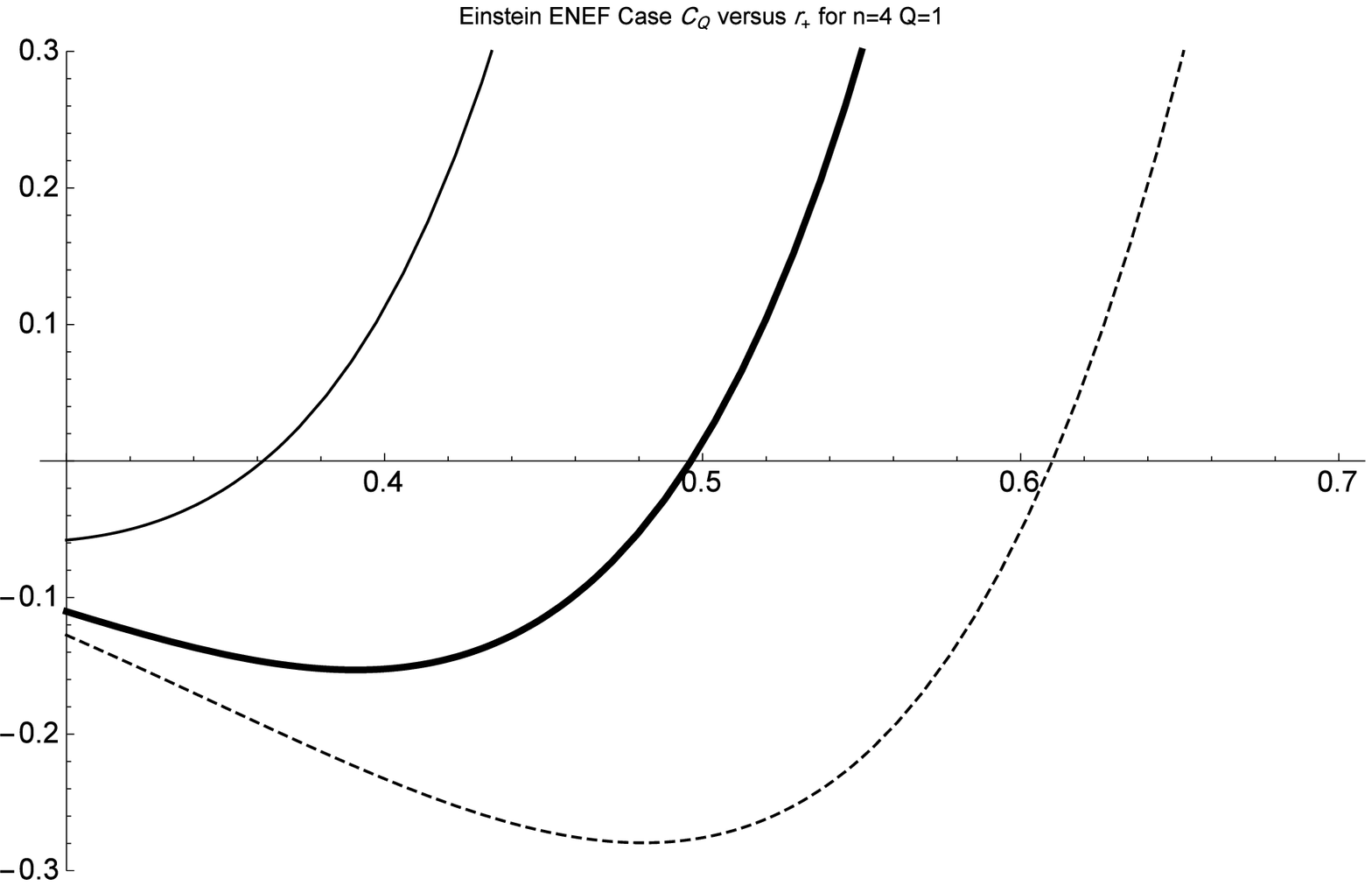} & \epsfxsize=7.5cm %
\epsffile{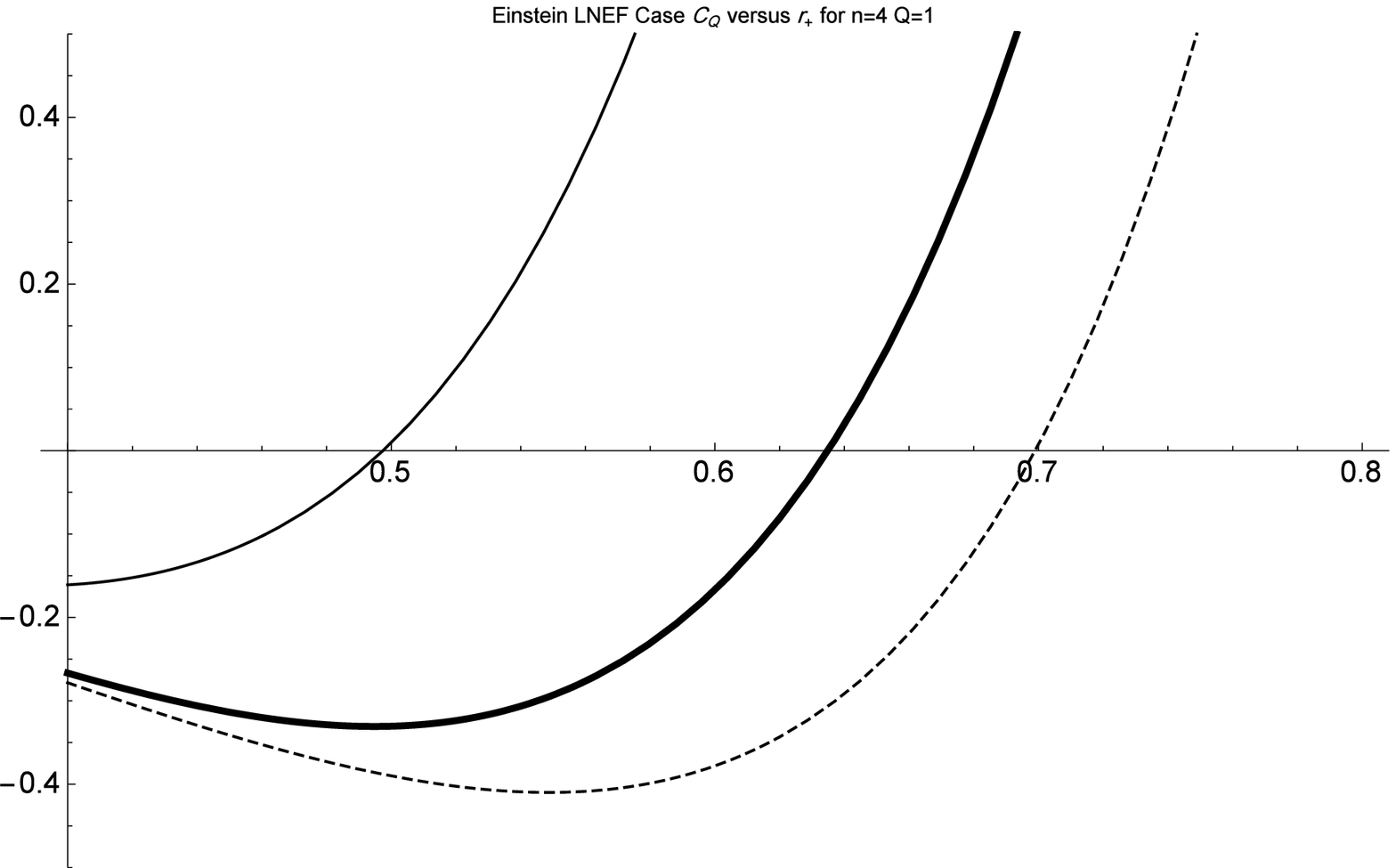}  \\
\epsfxsize=7.5cm \epsffile{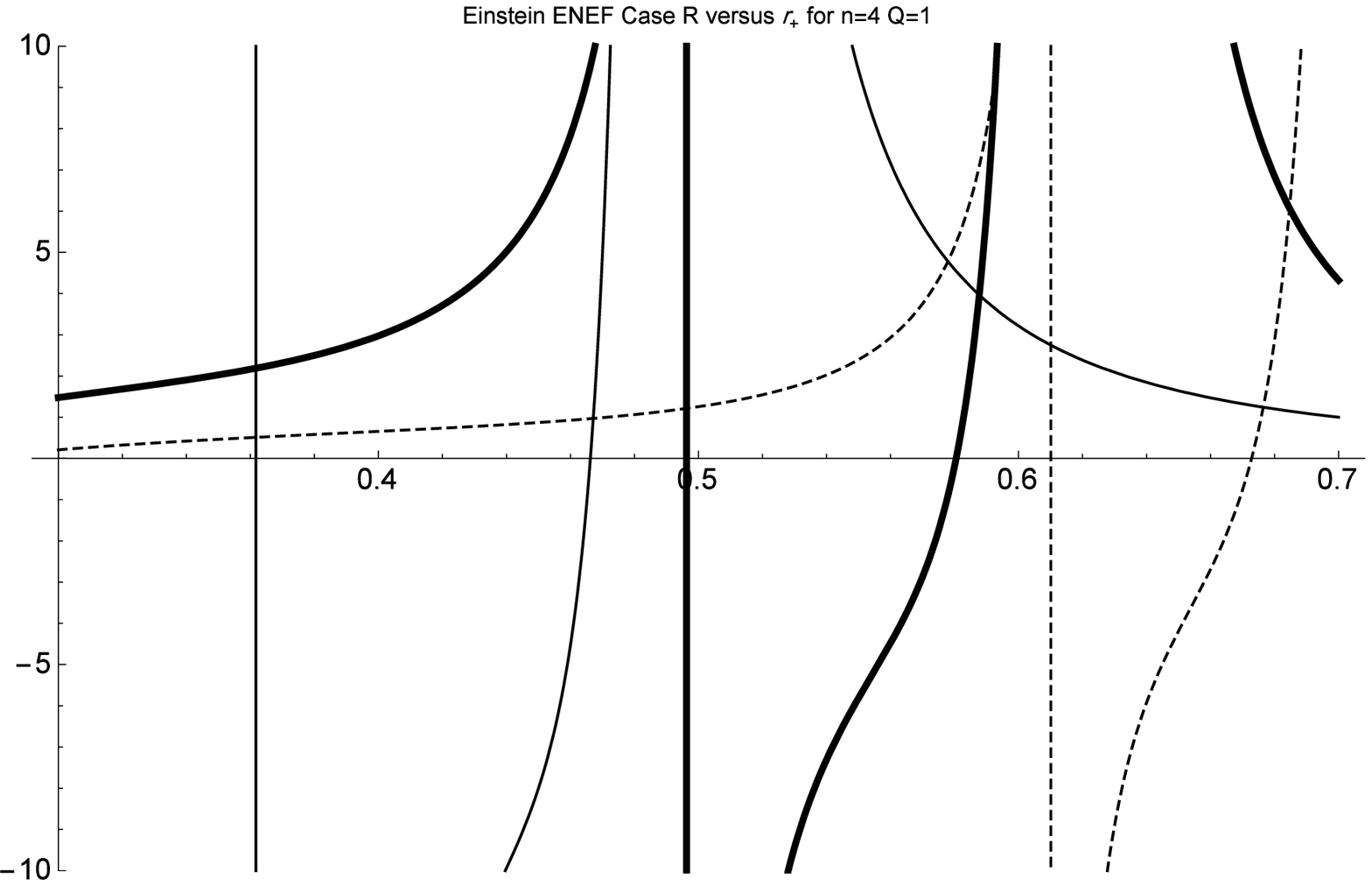} & \epsfxsize=7.5cm %
\epsffile{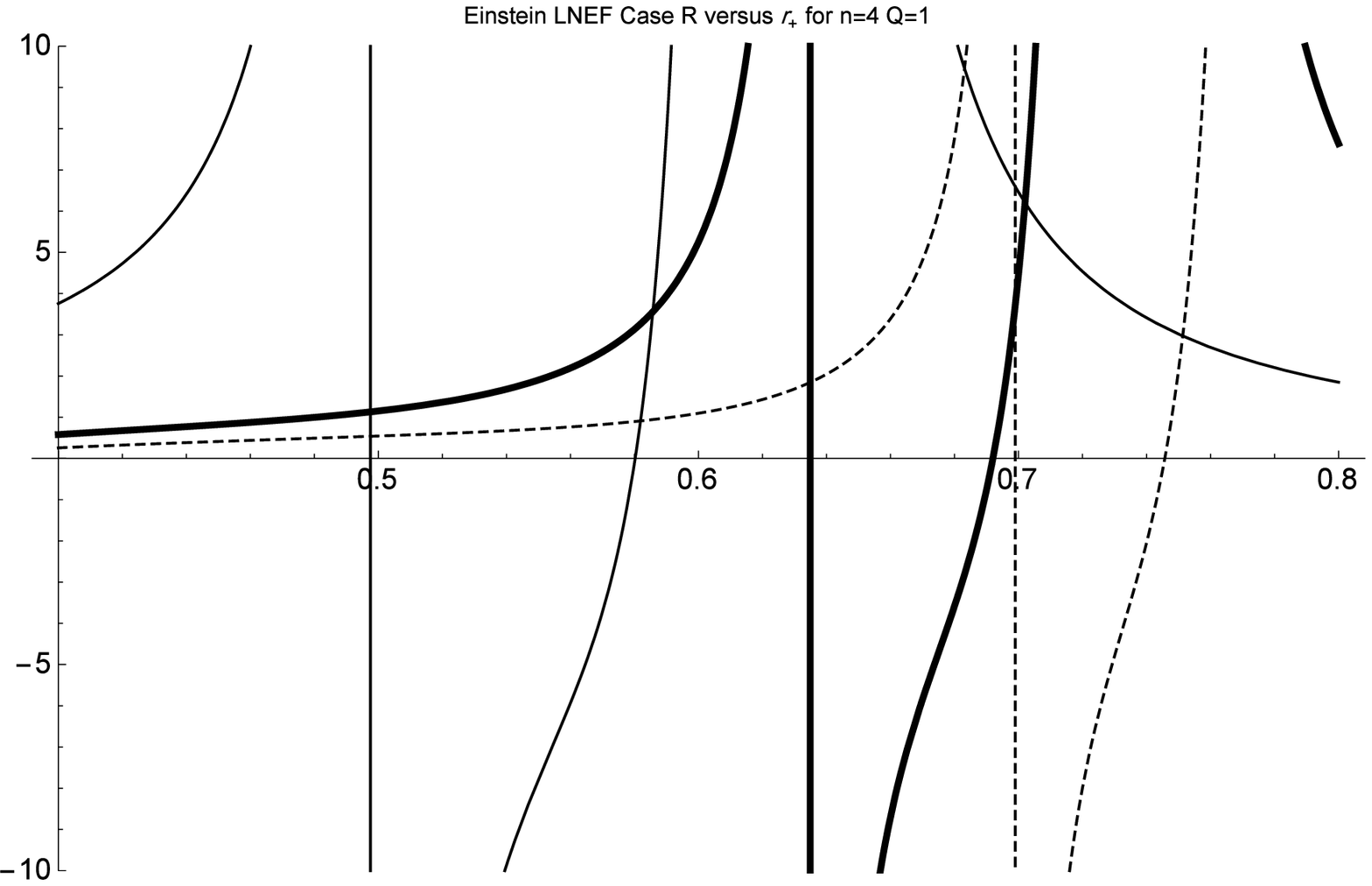}
\end{array}
$%
\caption{\textbf{"Einstein case: ENEF (left) and LNEF (right)
branches:"} Heat capacity (up) and Geometric Ricci scalar (down)
versus $r_{+}$ for $n=4$, $\Lambda=-1$, $Q=1$ and $\protect\beta
=0.5$ (solid line), $\protect\beta =1$ (bold line) and
$\protect\beta =2$ (dashed line). } \label{HEAT2}
\end{figure}

\begin{figure}[tbp]
$%
\begin{array}{cc}
\epsfxsize=7.5cm \epsffile{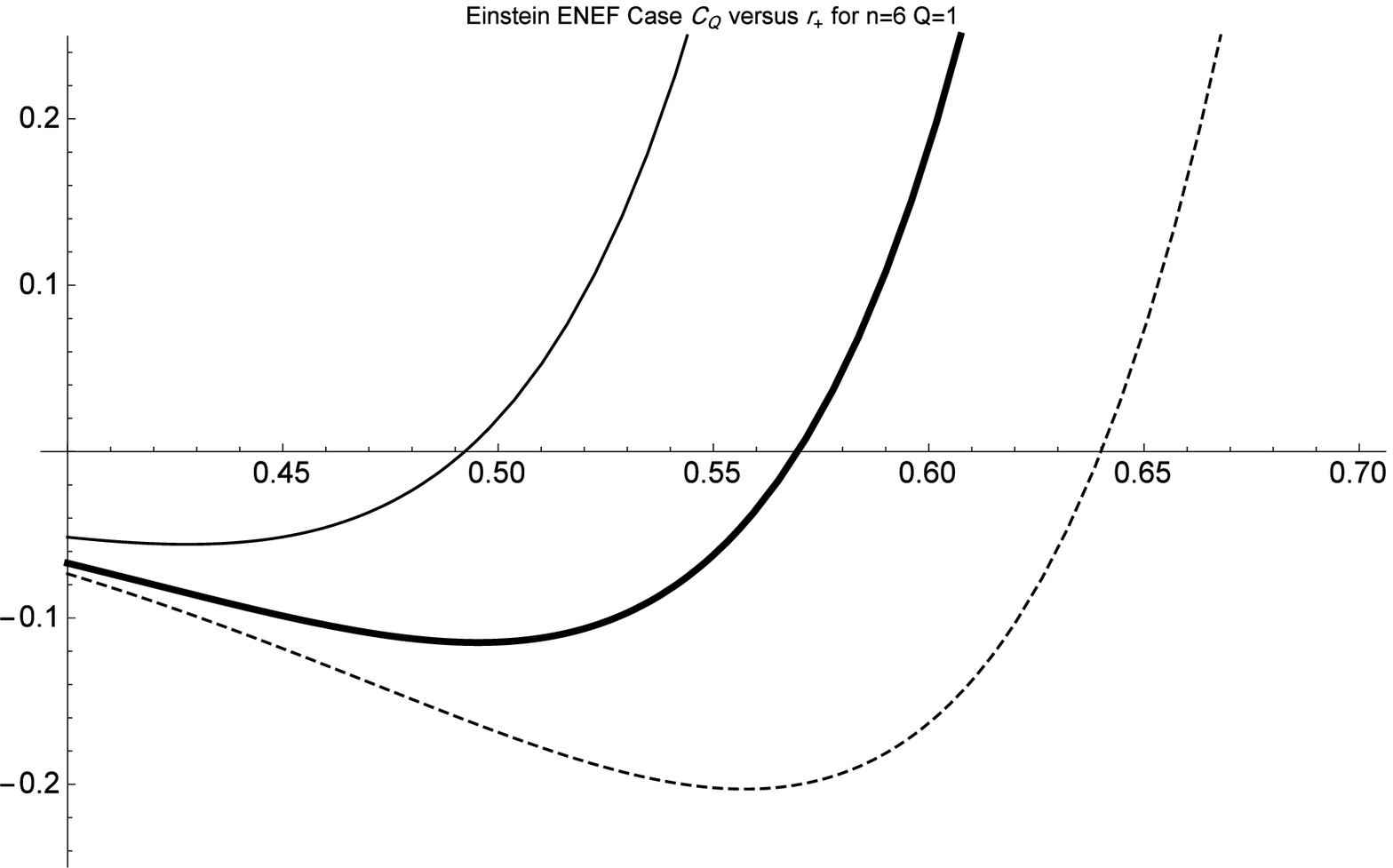} & \epsfxsize=7.5cm %
\epsffile{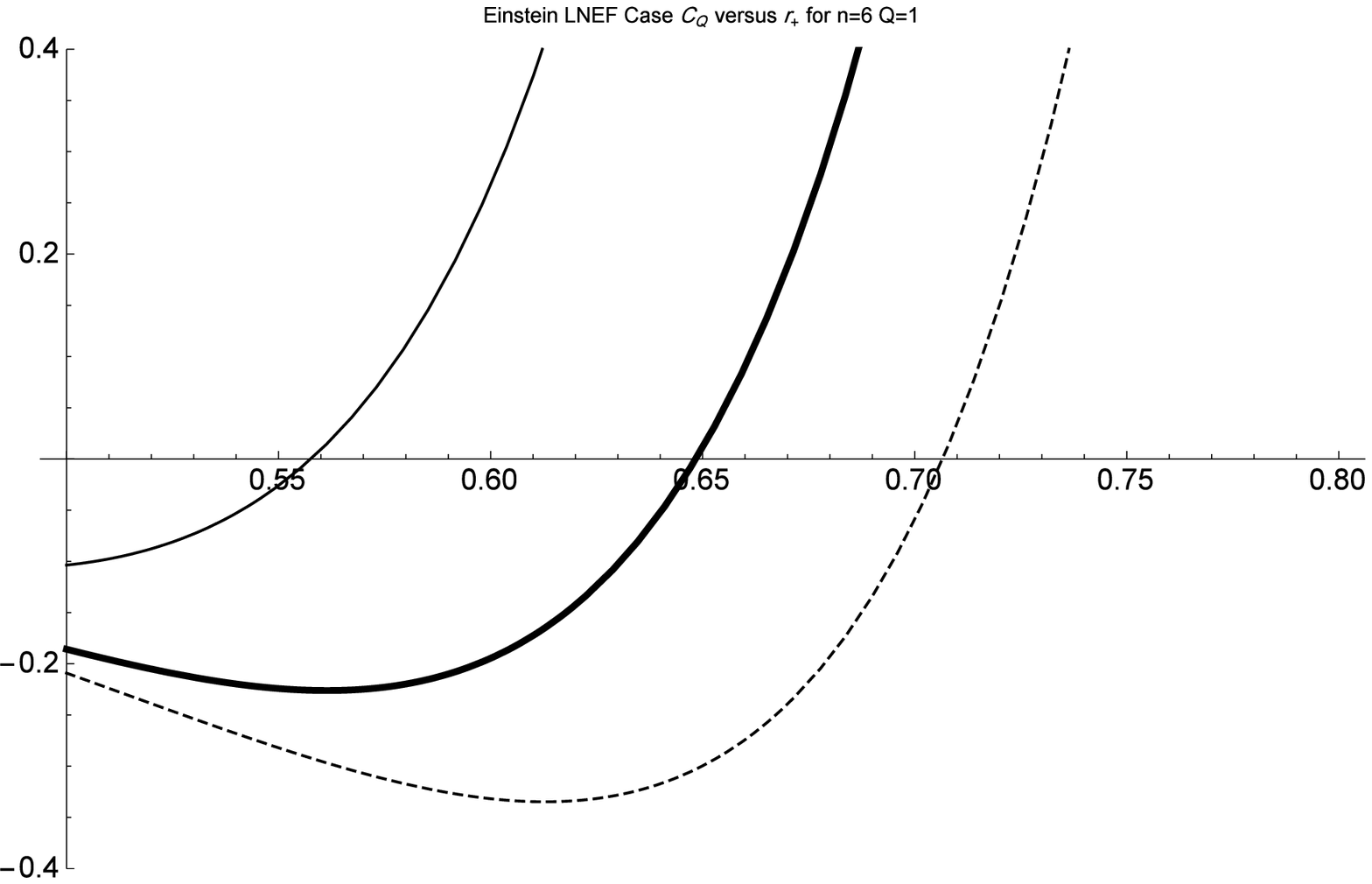}  \\
\epsfxsize=7.5cm \epsffile{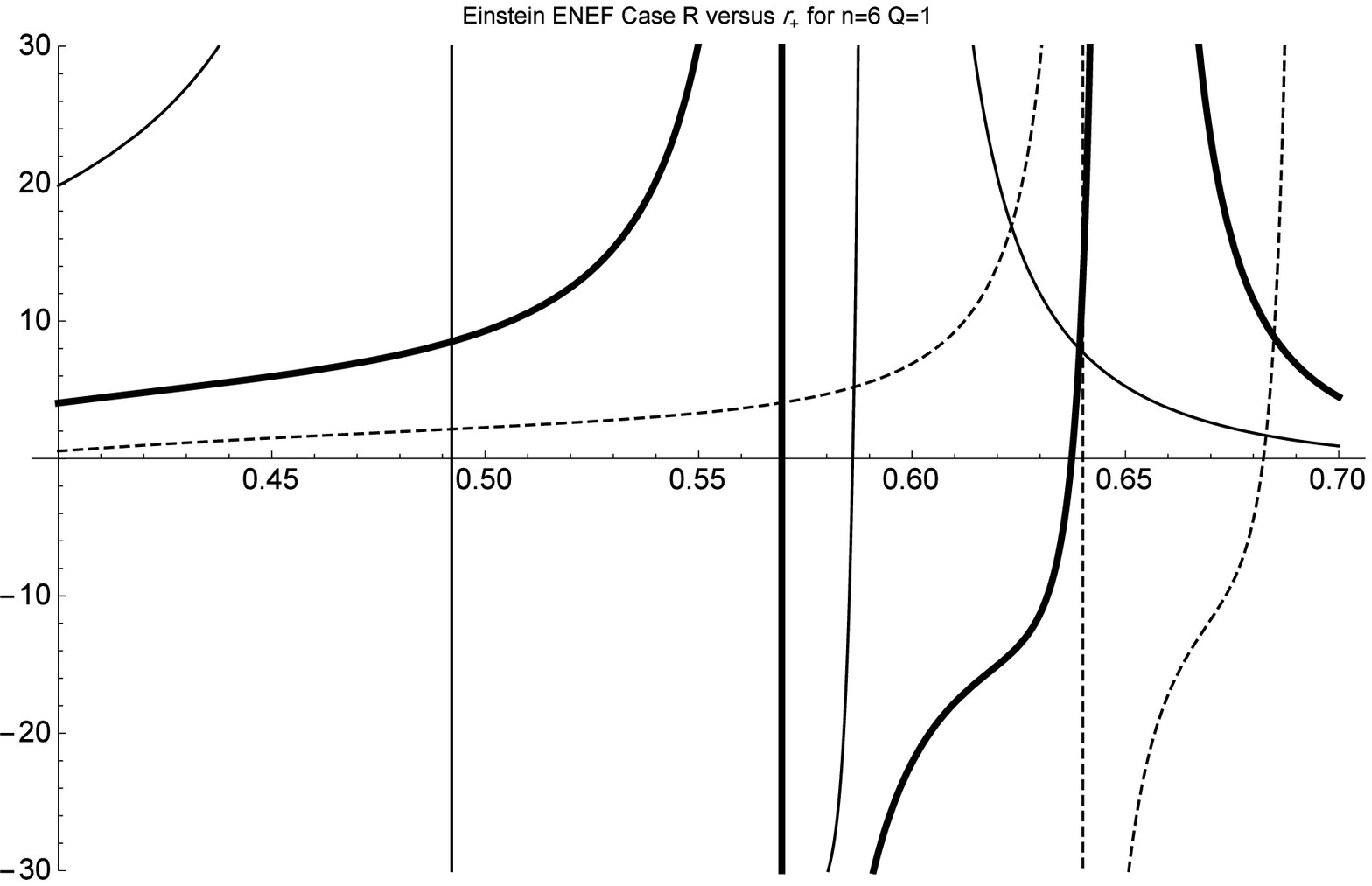} & \epsfxsize=7.5cm %
\epsffile{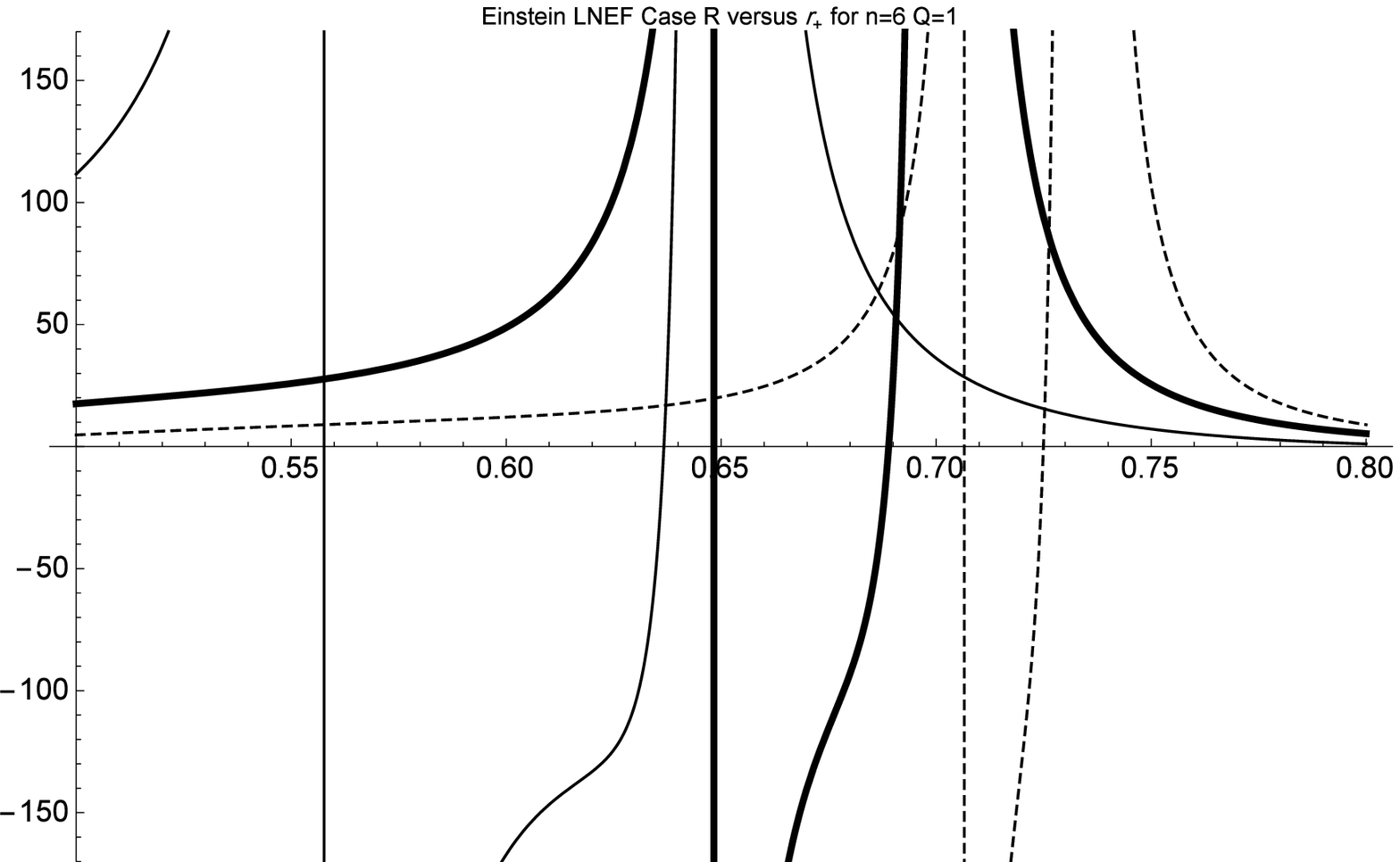}
\end{array}
$%
\caption{\textbf{"Einstein case: ENEF (left) and LNEF (right)
branches:"} Heat capacity (up) and Geometric Ricci scalar (down)
versus $r_{+}$ for $n=6$, $\Lambda=-1$, $Q=1$ and $\protect\beta
=0.5$ (solid line), $\protect\beta =1$ (bold line) and
$\protect\beta =2$ (dashed line). } \label{HEAT3}
\end{figure}

\begin{figure}[tbp]
$%
\begin{array}{cc}
\epsfxsize=7.5cm \epsffile{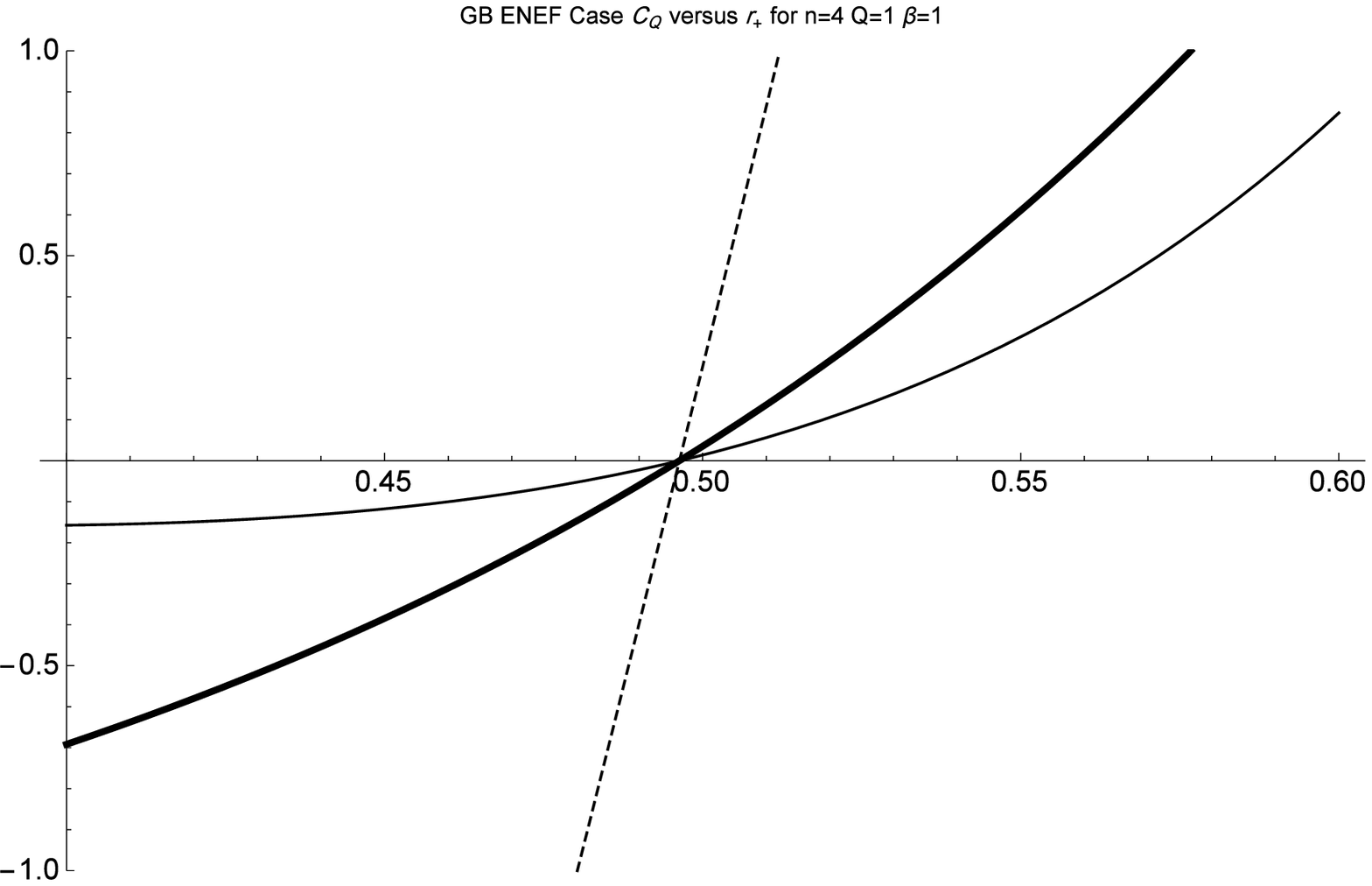} & \epsfxsize=7.5cm %
\epsffile{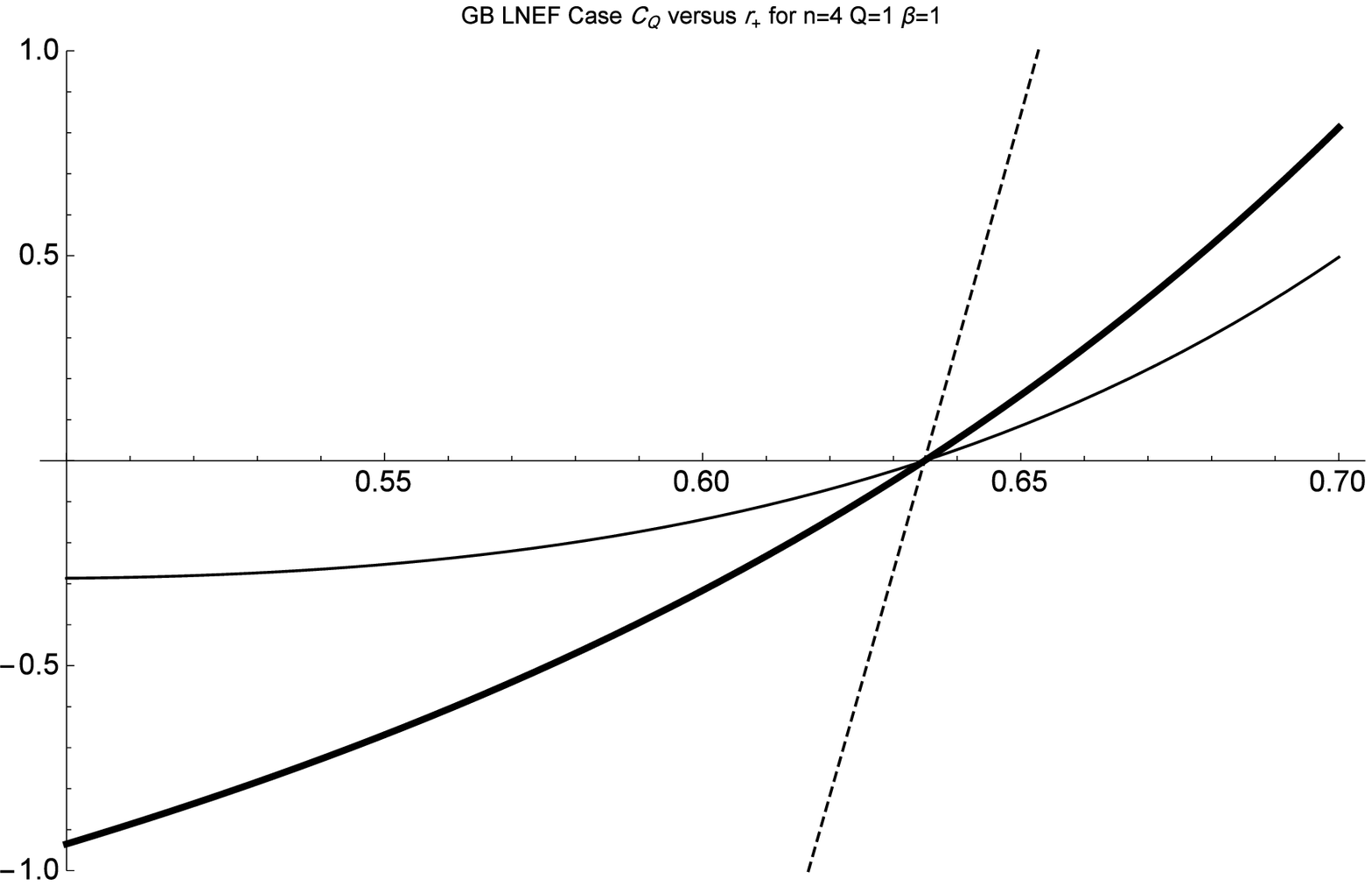}  \\
\epsfxsize=7.5cm \epsffile{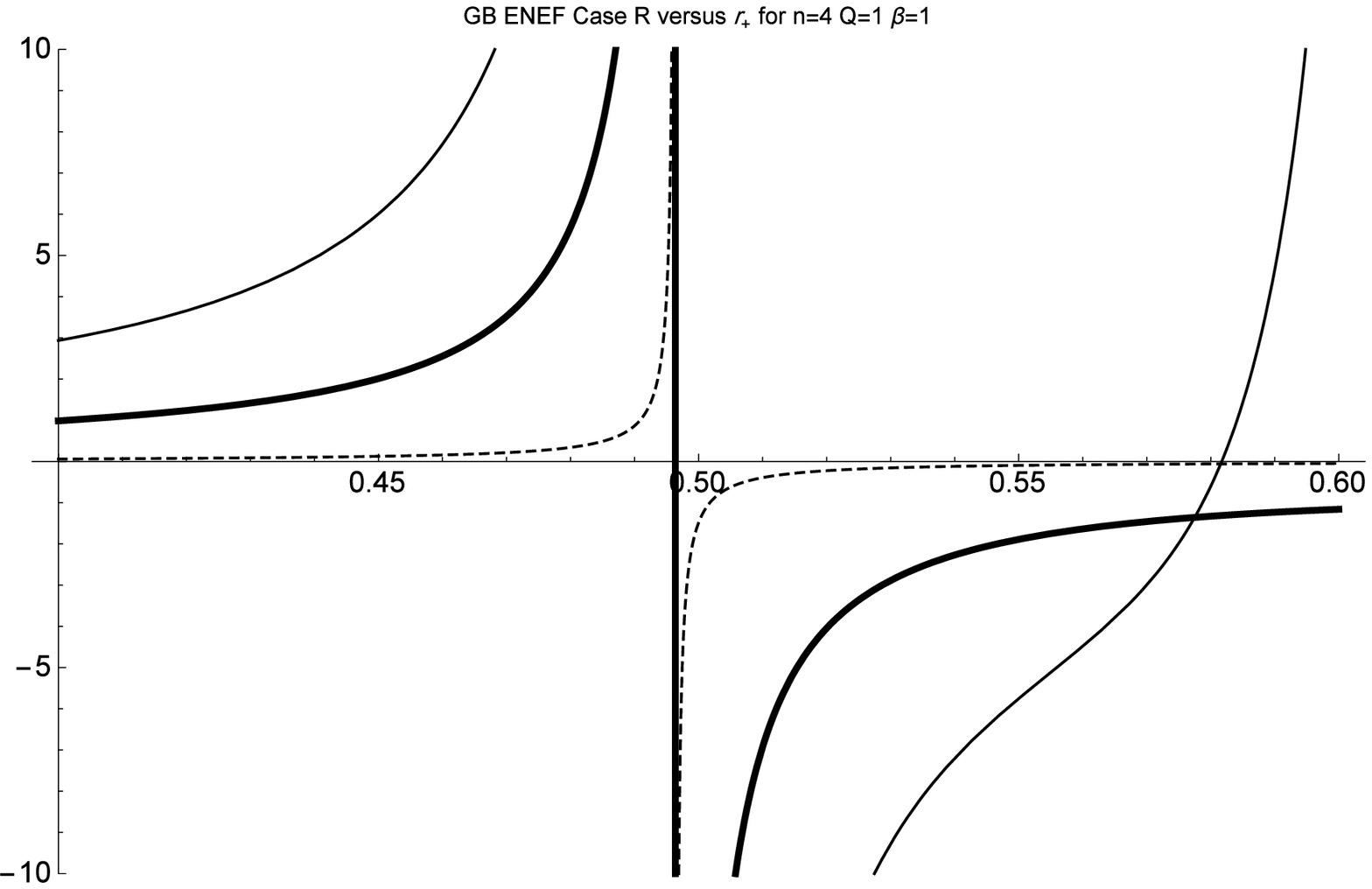} & \epsfxsize=7.5cm %
\epsffile{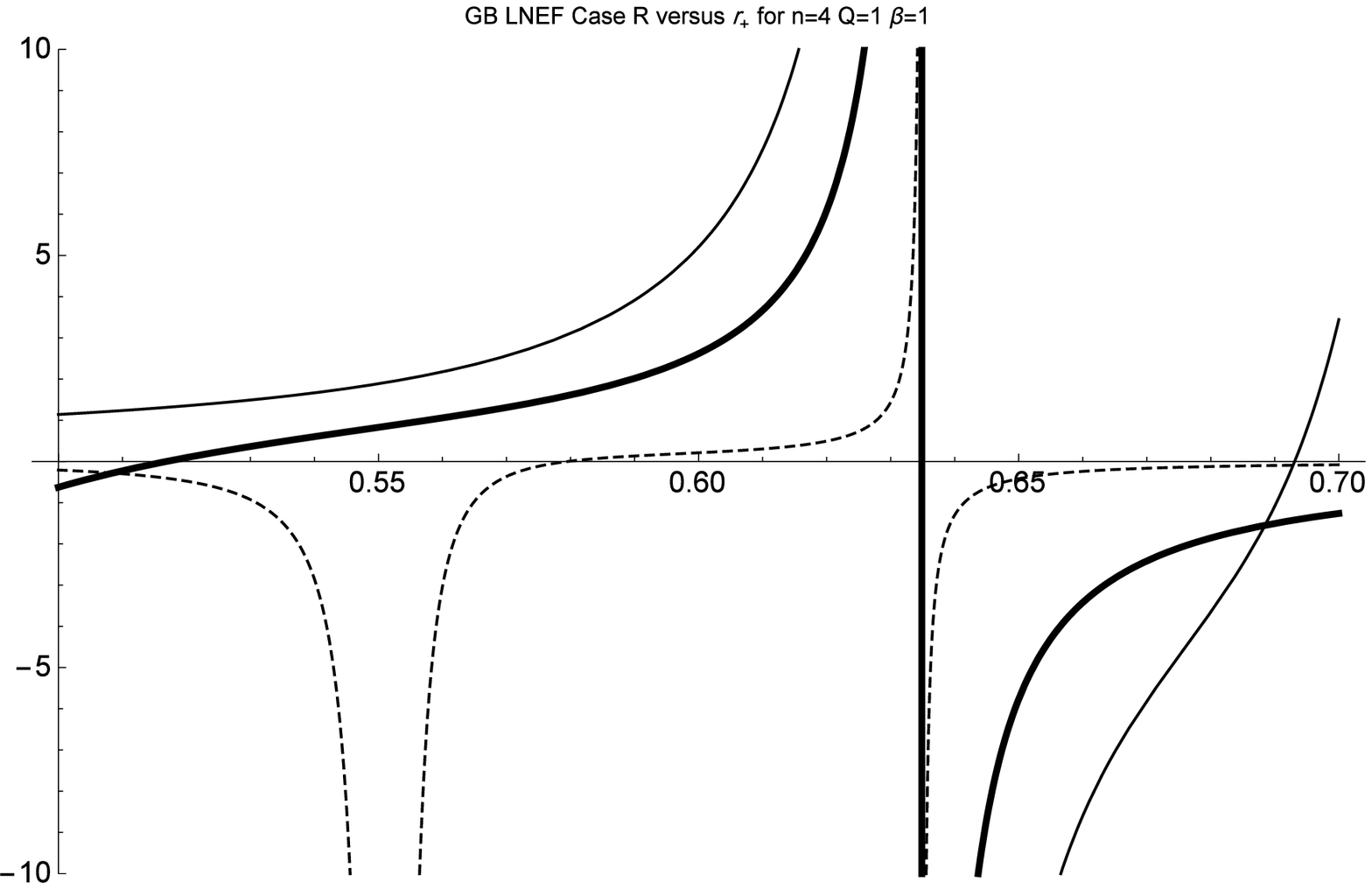}
\end{array}
$%
\caption{\textbf{"GB case: ENEF (left) and LNEF (right)
branches:"} Heat capacity (up) and Geometric Ricci scalar (down)
versus $r_{+}$ for $n=4$, $\Lambda=-1$, $Q=1$, $\protect\beta=1$
and $\protect \alpha =0.001$ (solid line), $\protect\alpha =0.1$
(bold line) and $\protect\alpha =1$ (dashed line). } \label{HEAT4}
\end{figure}

\begin{figure}[tbp]
$%
\begin{array}{cc}
\epsfxsize=7.5cm \epsffile{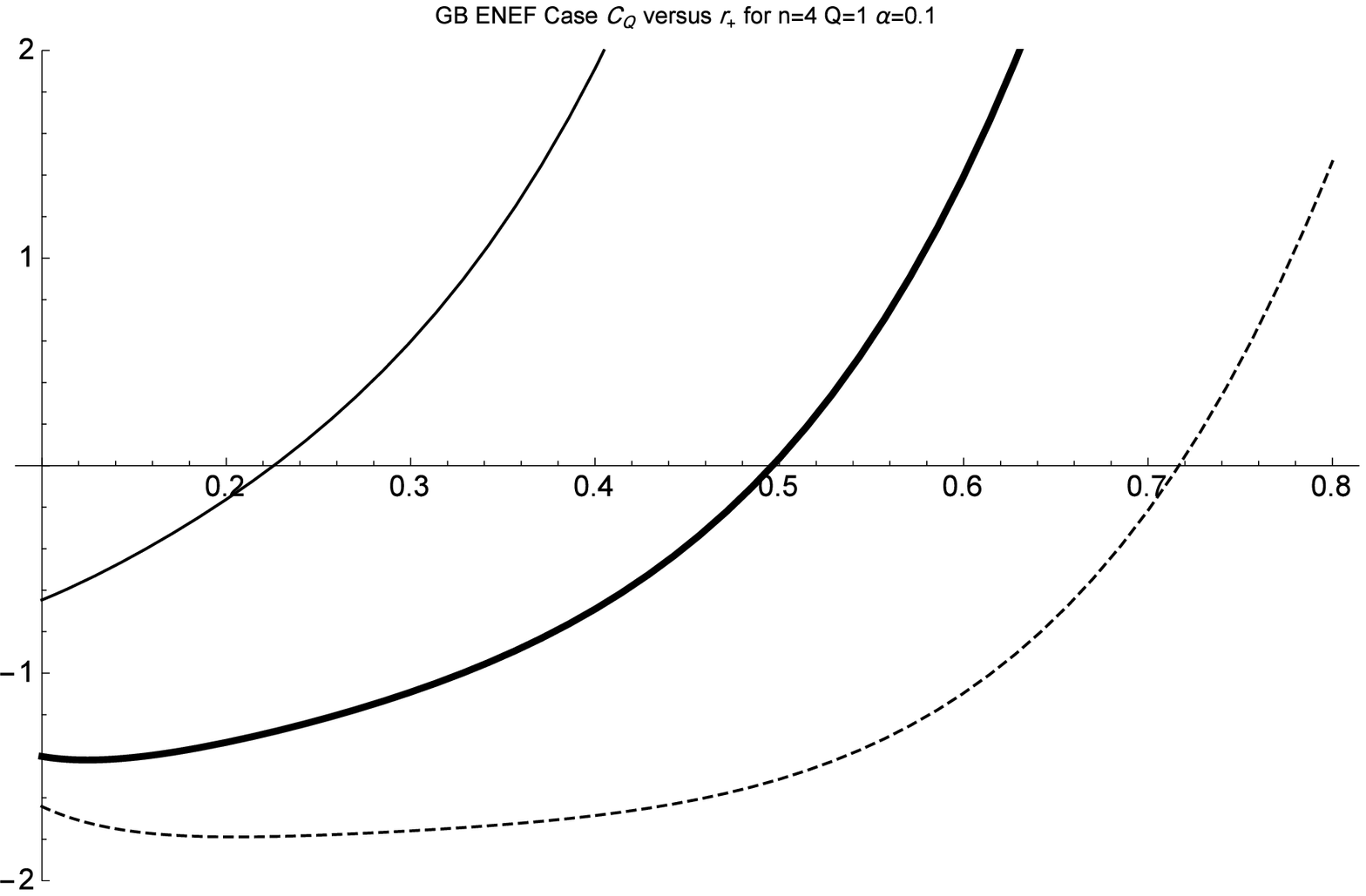} & \epsfxsize=7.5cm %
\epsffile{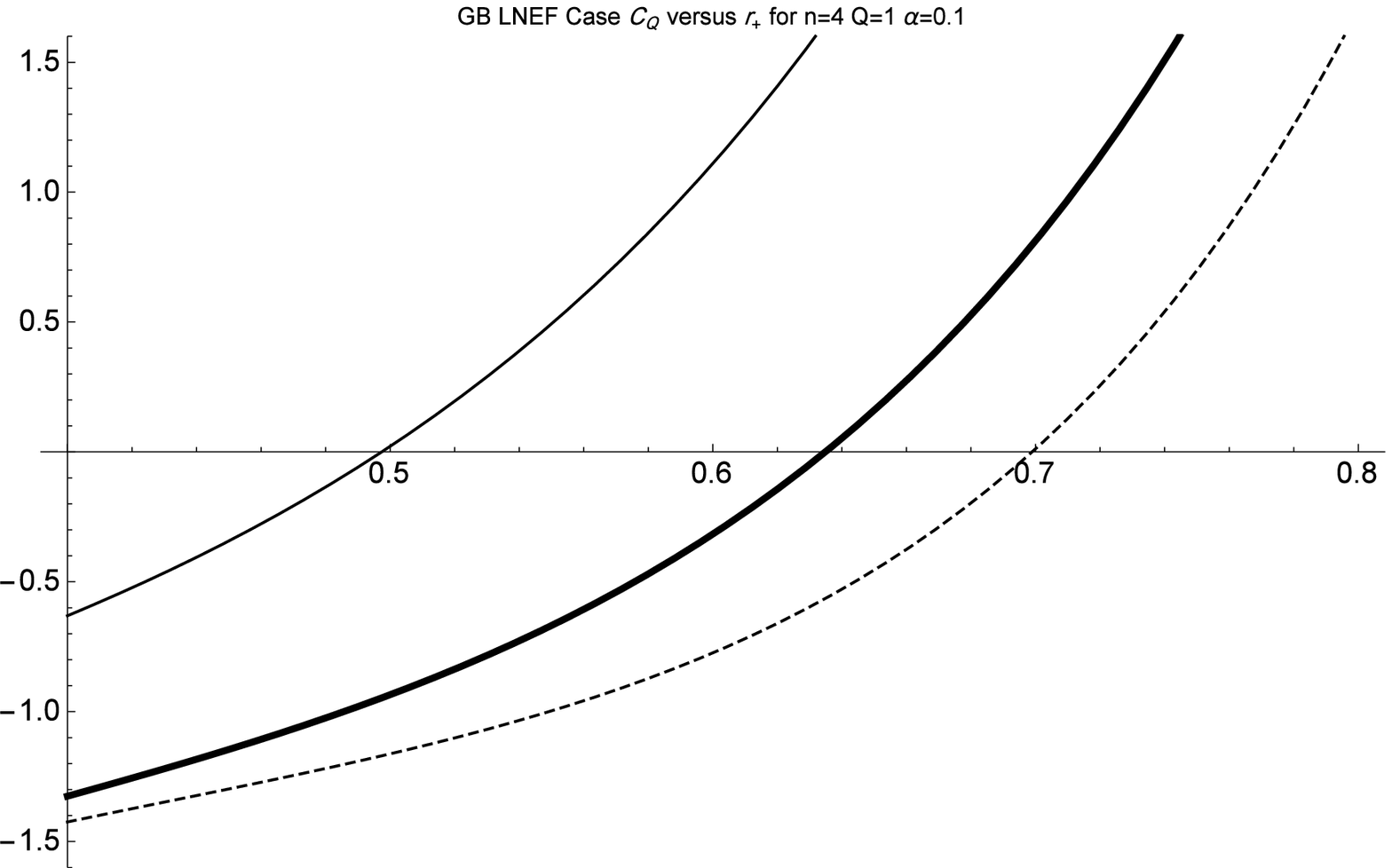}   \\
\epsfxsize=7.5cm \epsffile{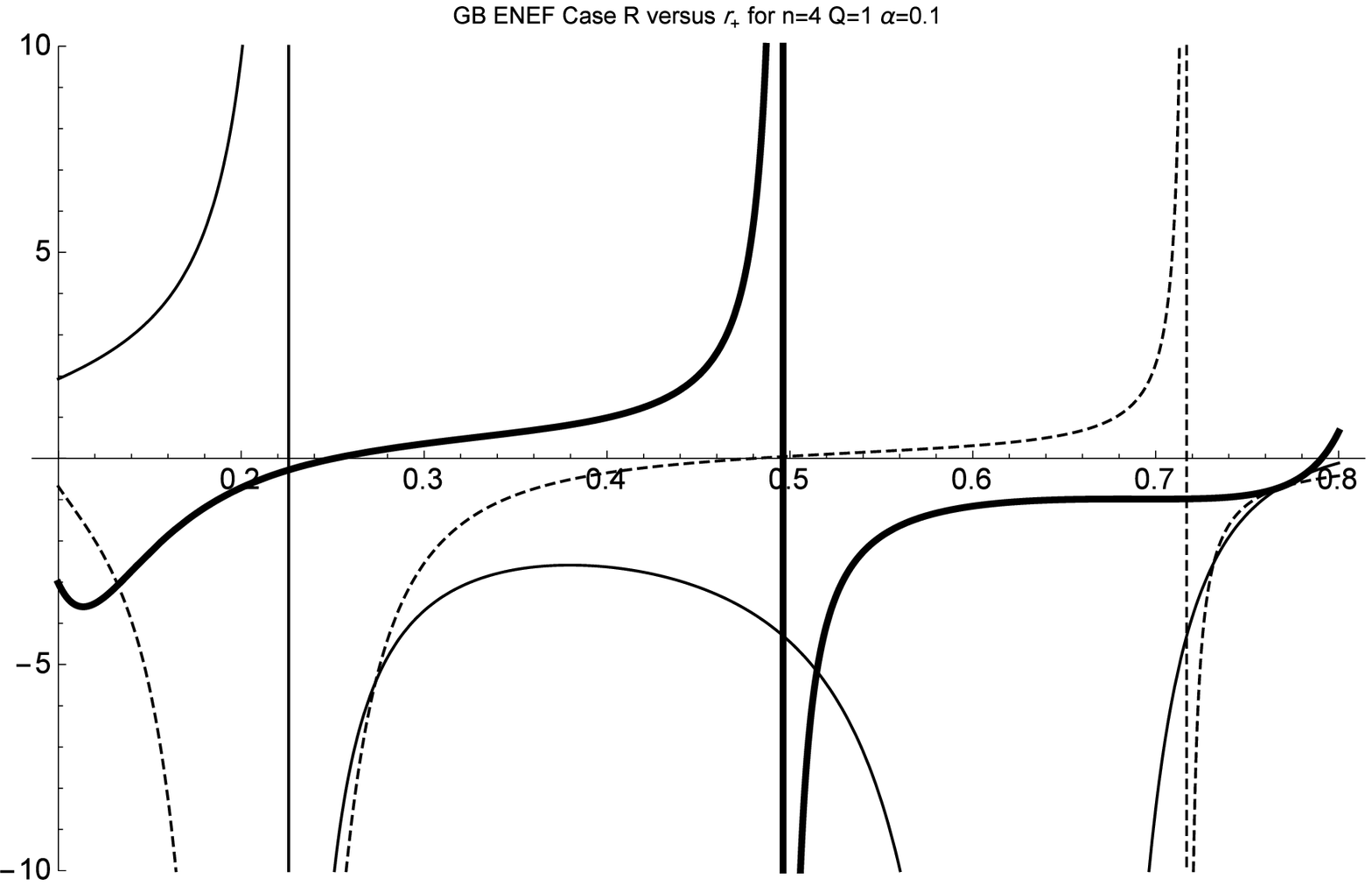} & \epsfxsize=7.5cm %
\epsffile{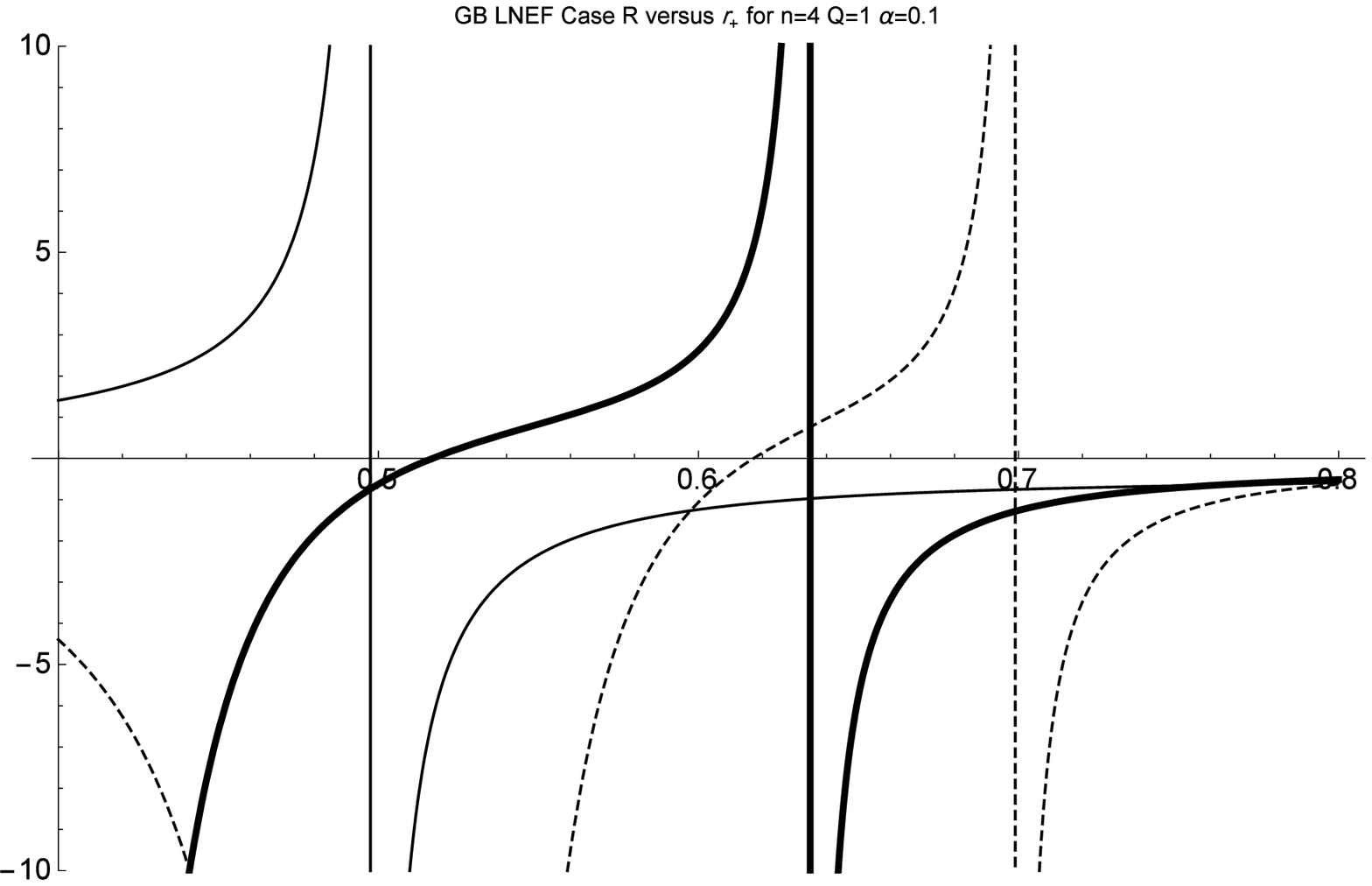}
\end{array}
$%
\caption{\textbf{"GB case: ENEF (left) and LNEF (right)
branches:"} Heat capacity (up) and Geometric Ricci scalar (down)
versus $r_{+}$ for $n=4$, $\Lambda=-1$, $Q=1$,
$\protect\alpha=0.1$ and $\protect \beta =0.5$ (solid line),
$\protect\beta =1$ (bold line) and $\protect \beta =2$ (dashed
line). } \label{HEAT5}
\end{figure}

\begin{figure}[tbp]
$%
\begin{array}{cc}
\epsfxsize=7.5cm \epsffile{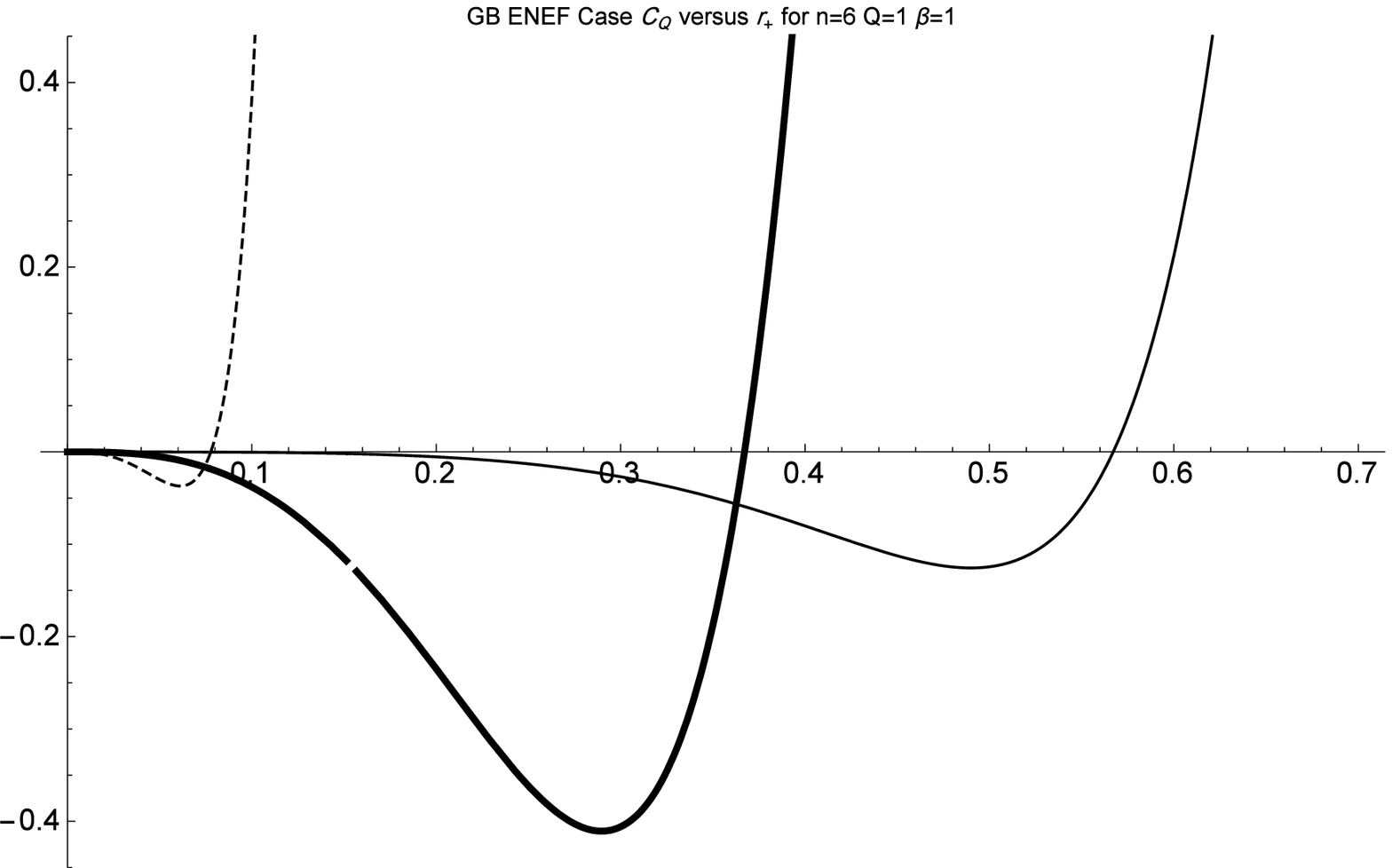} & \epsfxsize=7.5cm %
\epsffile{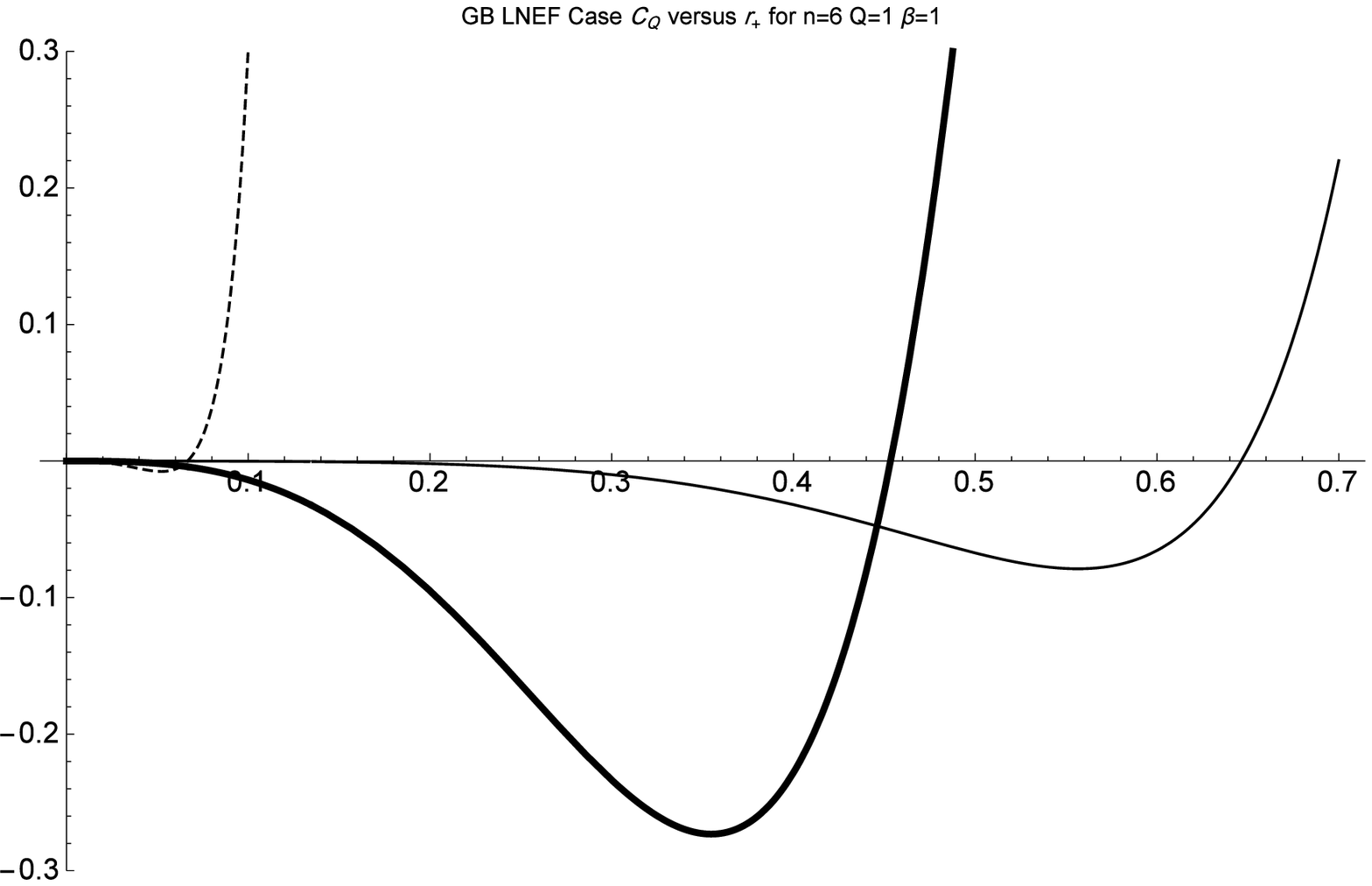} \\
\epsfxsize=7.5cm \epsffile{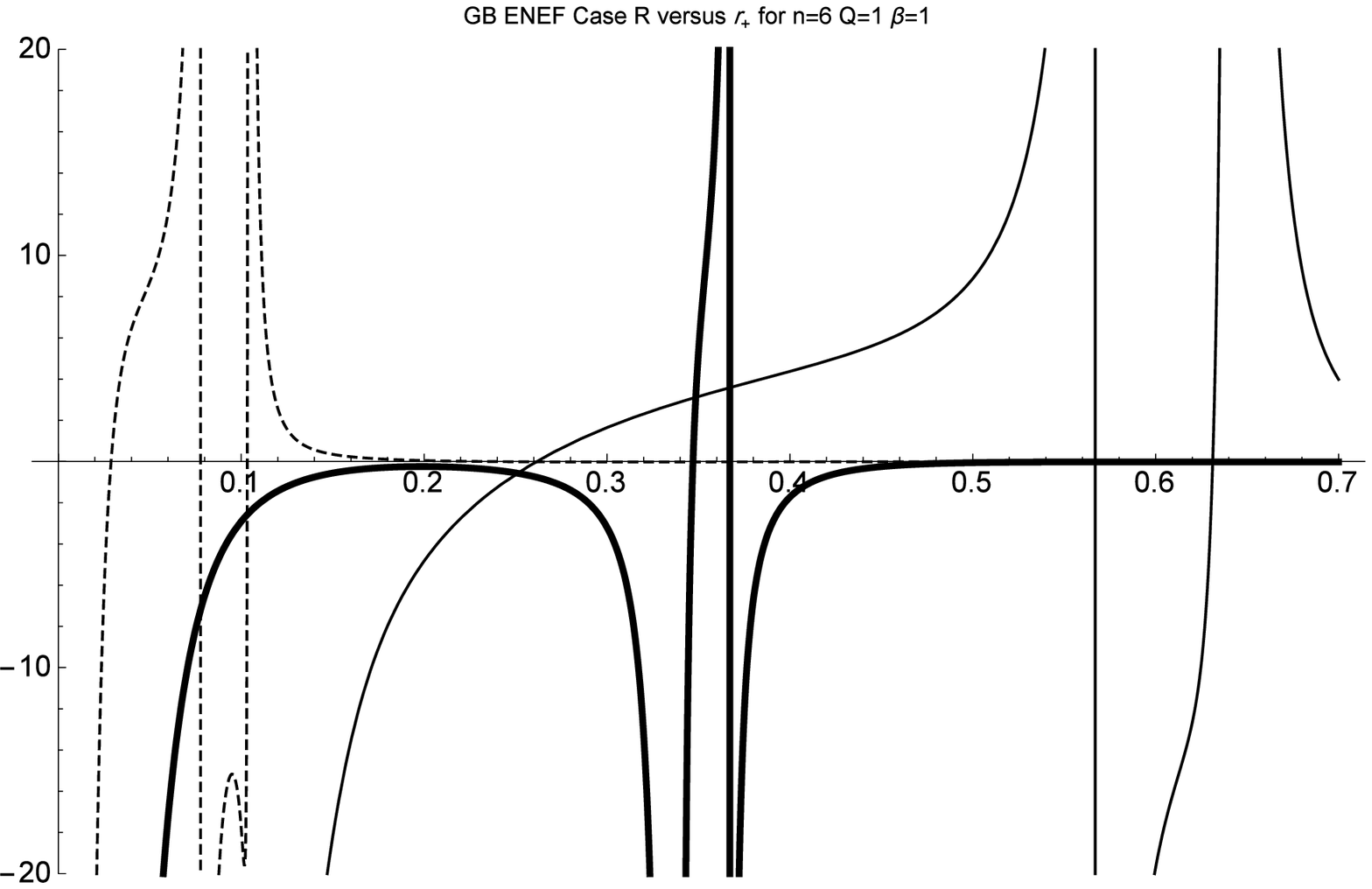} & \epsfxsize=7.5cm %
\epsffile{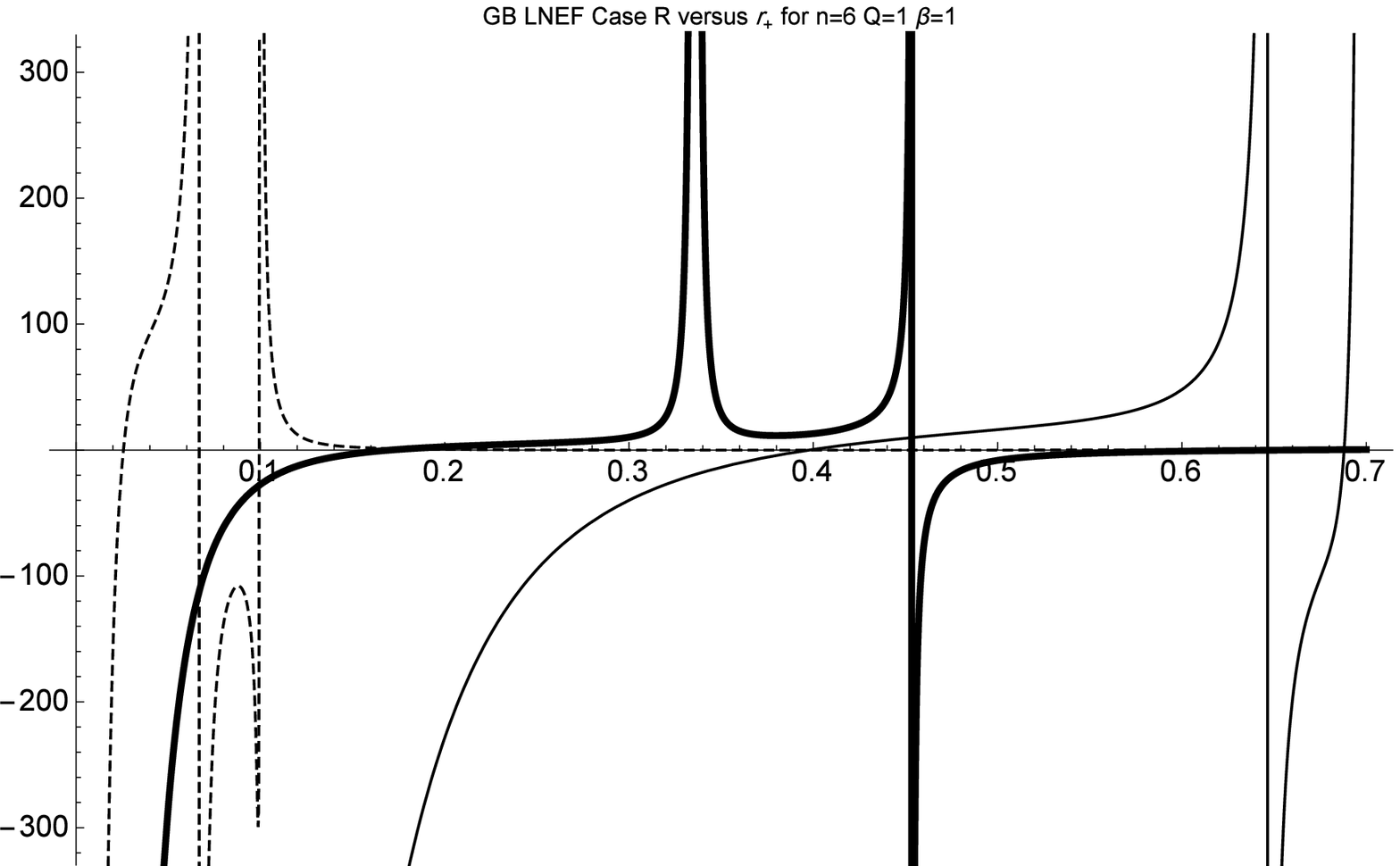}
\end{array}
$%
\caption{\textbf{"GB case: ENEF (left) and LNEF (right)
branches:"} Heat capacity (up) and Geometric Ricci scalar (down)
versus $r_{+}$ for $n=6$, $\Lambda=-1$, $Q=1$, $\protect\beta=1$
and $\protect \alpha =0.001$ (solid line), $\protect\alpha =0.1$
(bold line) and $\protect\alpha =1$ (dashed line). } \label{HEAT6}
\end{figure}

\begin{figure}[tbp]
$%
\begin{array}{cc}
\epsfxsize=7.5cm \epsffile{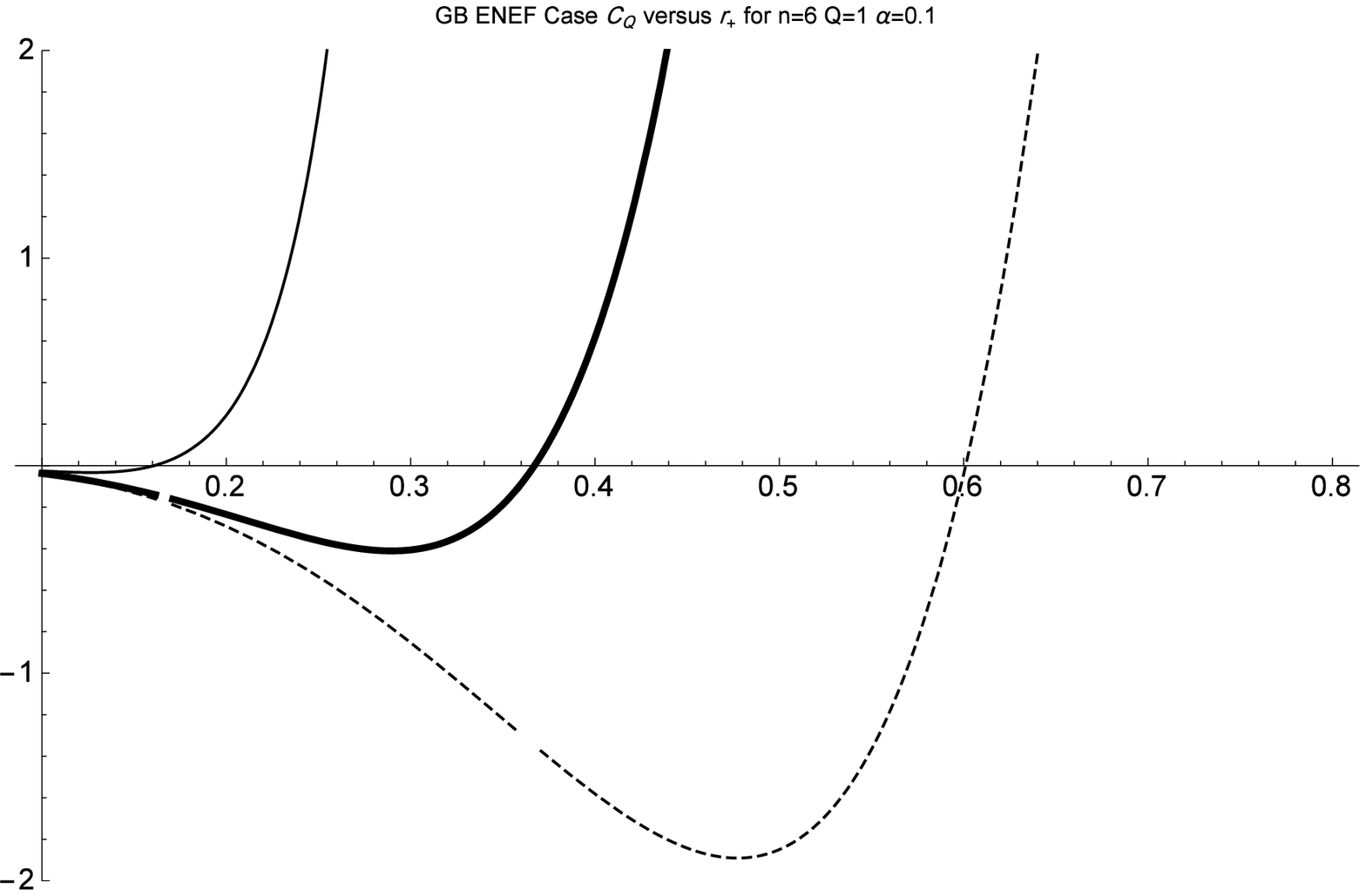} & \epsfxsize=7.5cm %
\epsffile{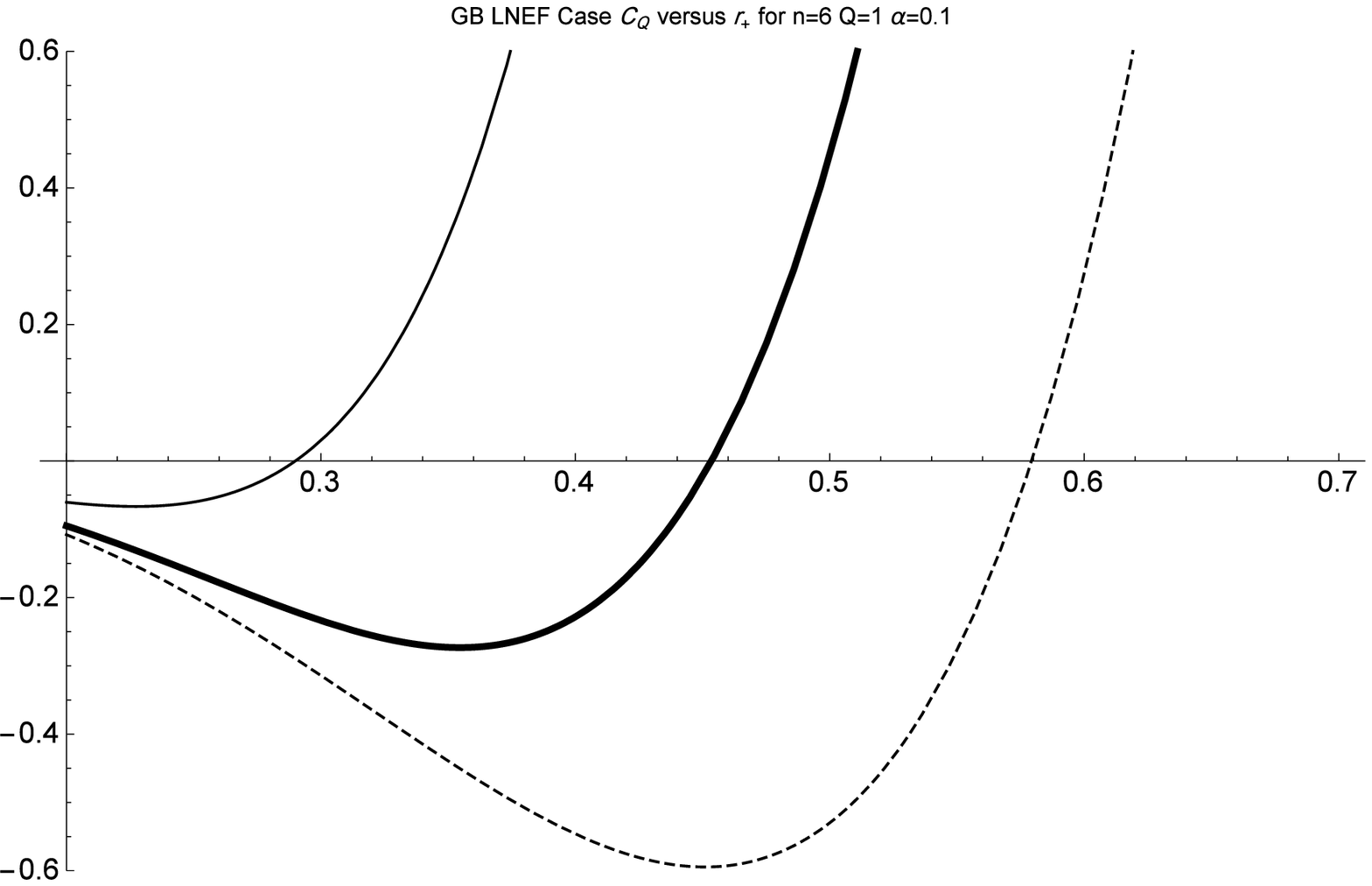}   \\
\epsfxsize=7.5cm \epsffile{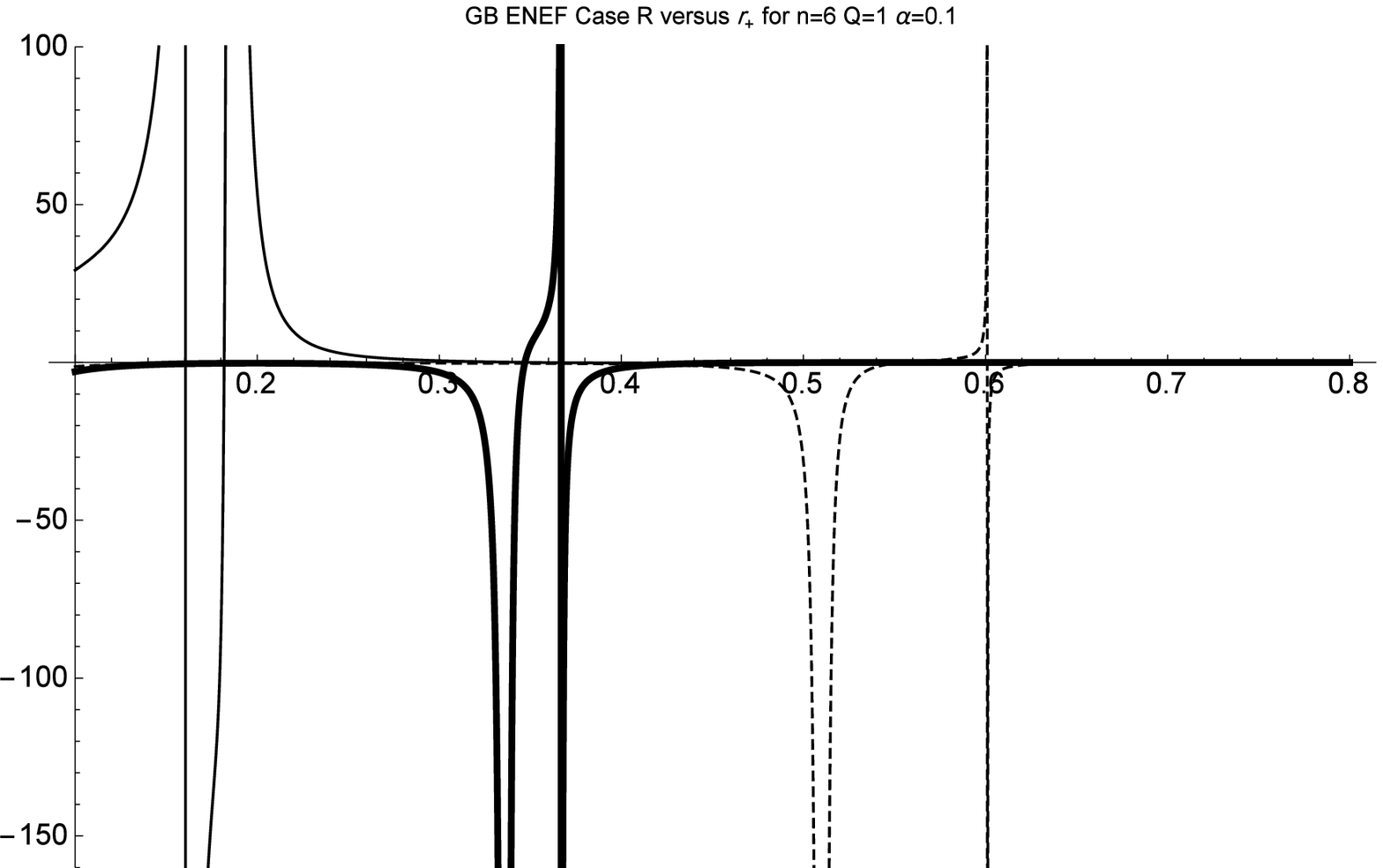} & \epsfxsize=7.5cm %
\epsffile{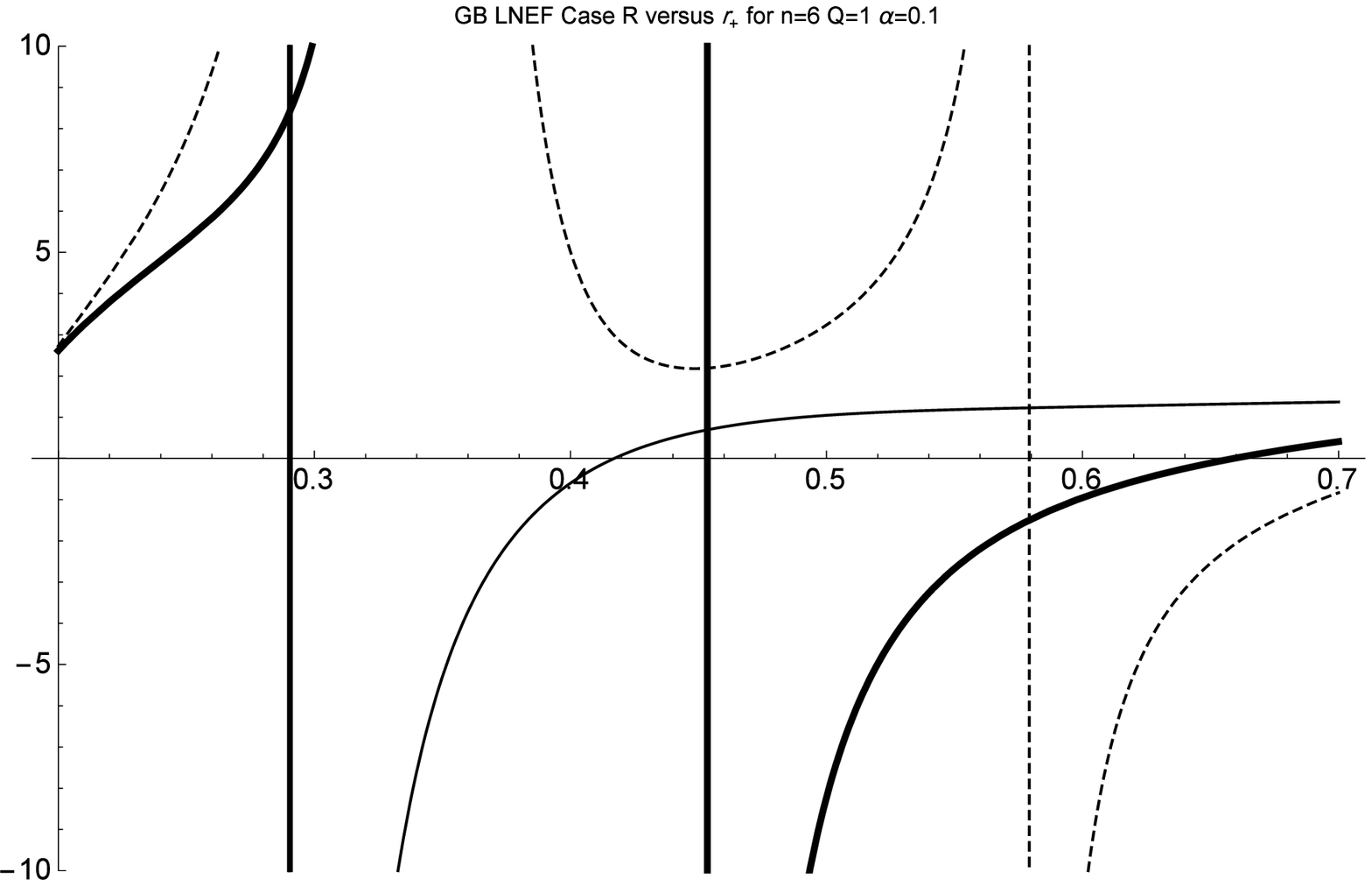}
\end{array}
$%
\caption{\textbf{"GB case: ENEF (left) and LNEF (right)
branches:"} Heat capacity (up) and Geometric Ricci scalar (down)
versus $r_{+}$ for $n=6$, $\Lambda=-1$, $Q=1$,
$\protect\alpha=0.1$ and $\protect \beta =0.5$ (solid line),
$\protect\beta =1$ (bold line) and $\protect \beta =2$ (dashed
line). } \label{HEAT7}
\end{figure}

\begin{figure}[tbp]
$%
\begin{array}{cc}
\epsfxsize=7.5cm \epsffile{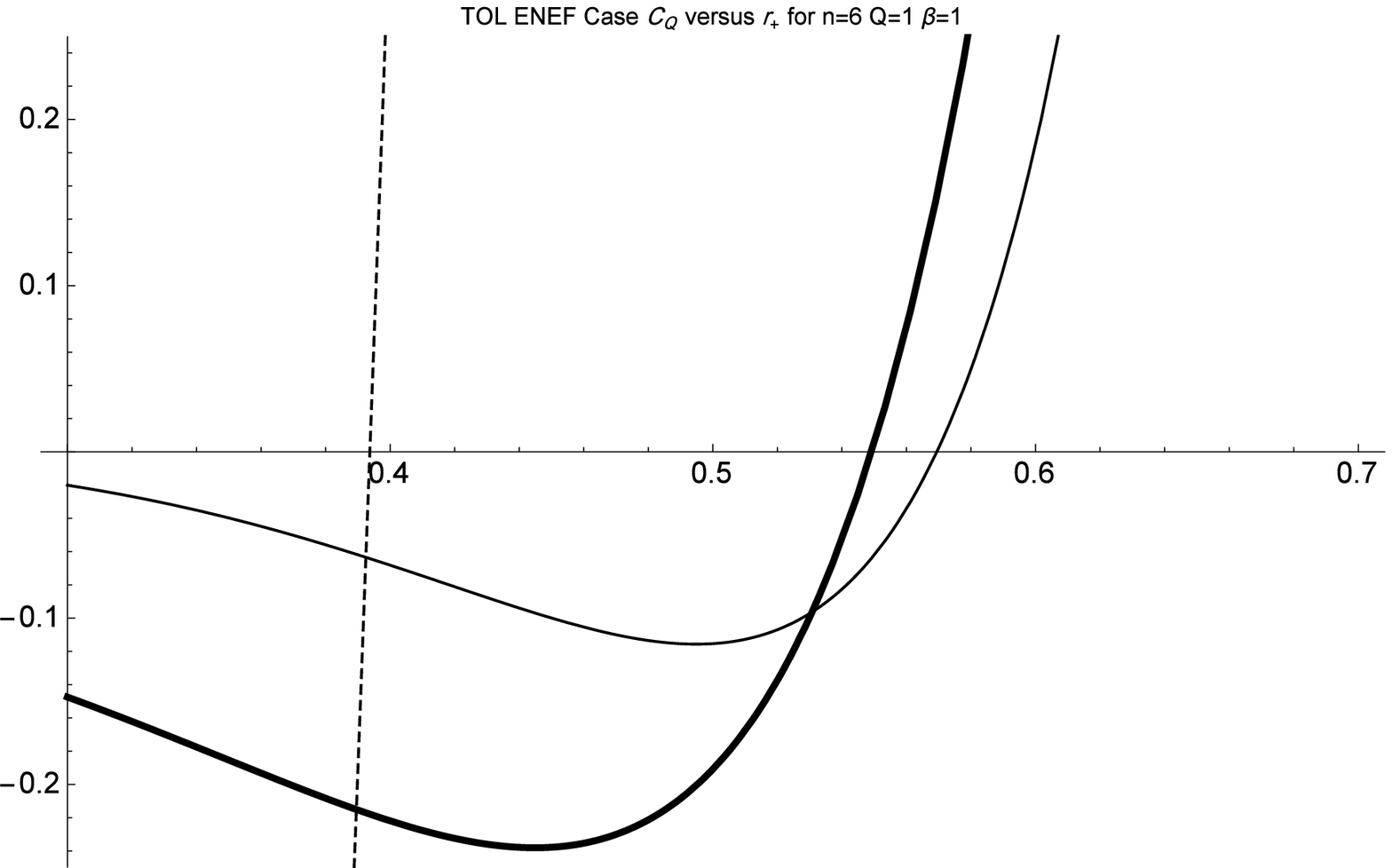} & \epsfxsize=7.5cm %
\epsffile{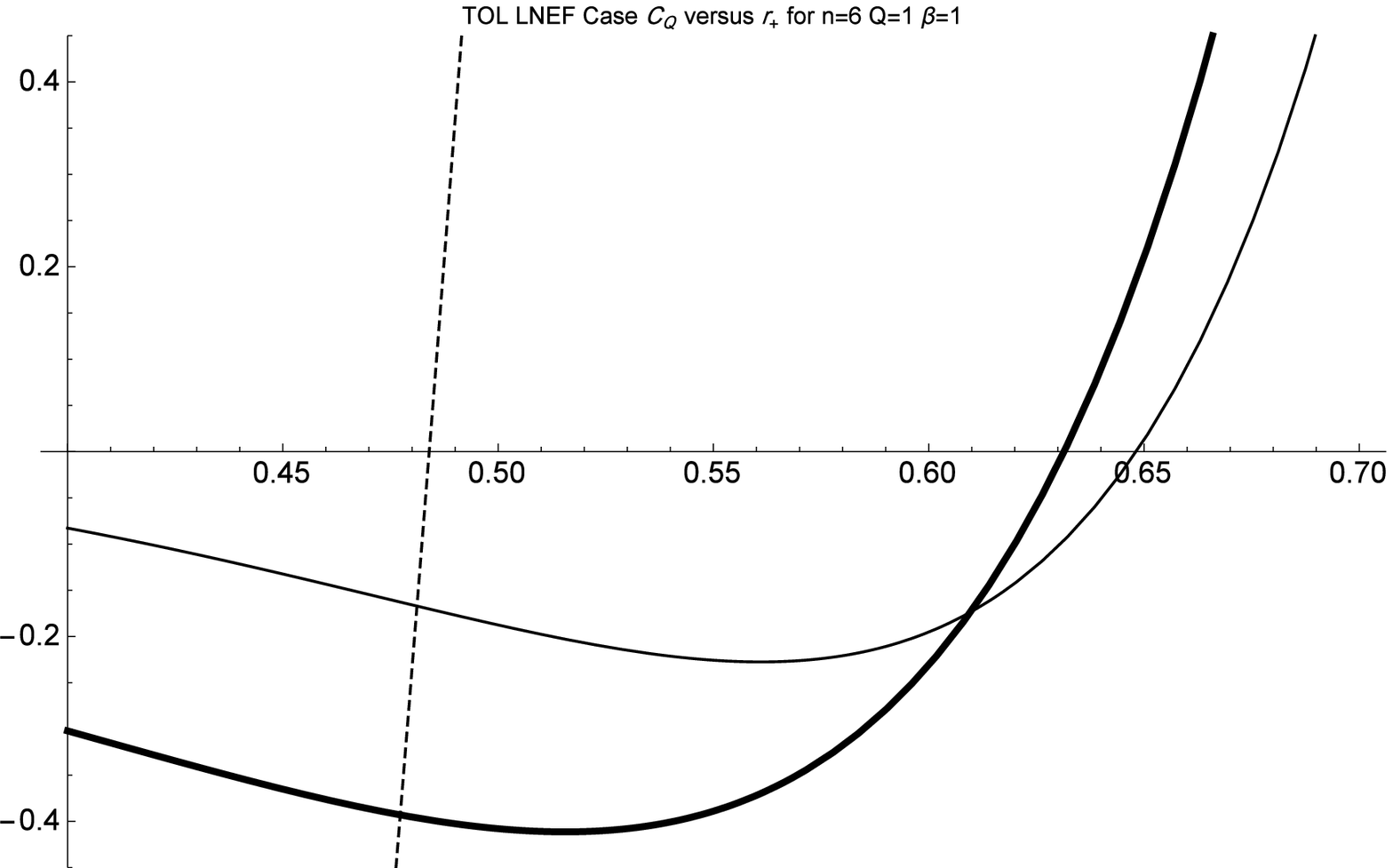}   \\
\epsfxsize=7.5cm \epsffile{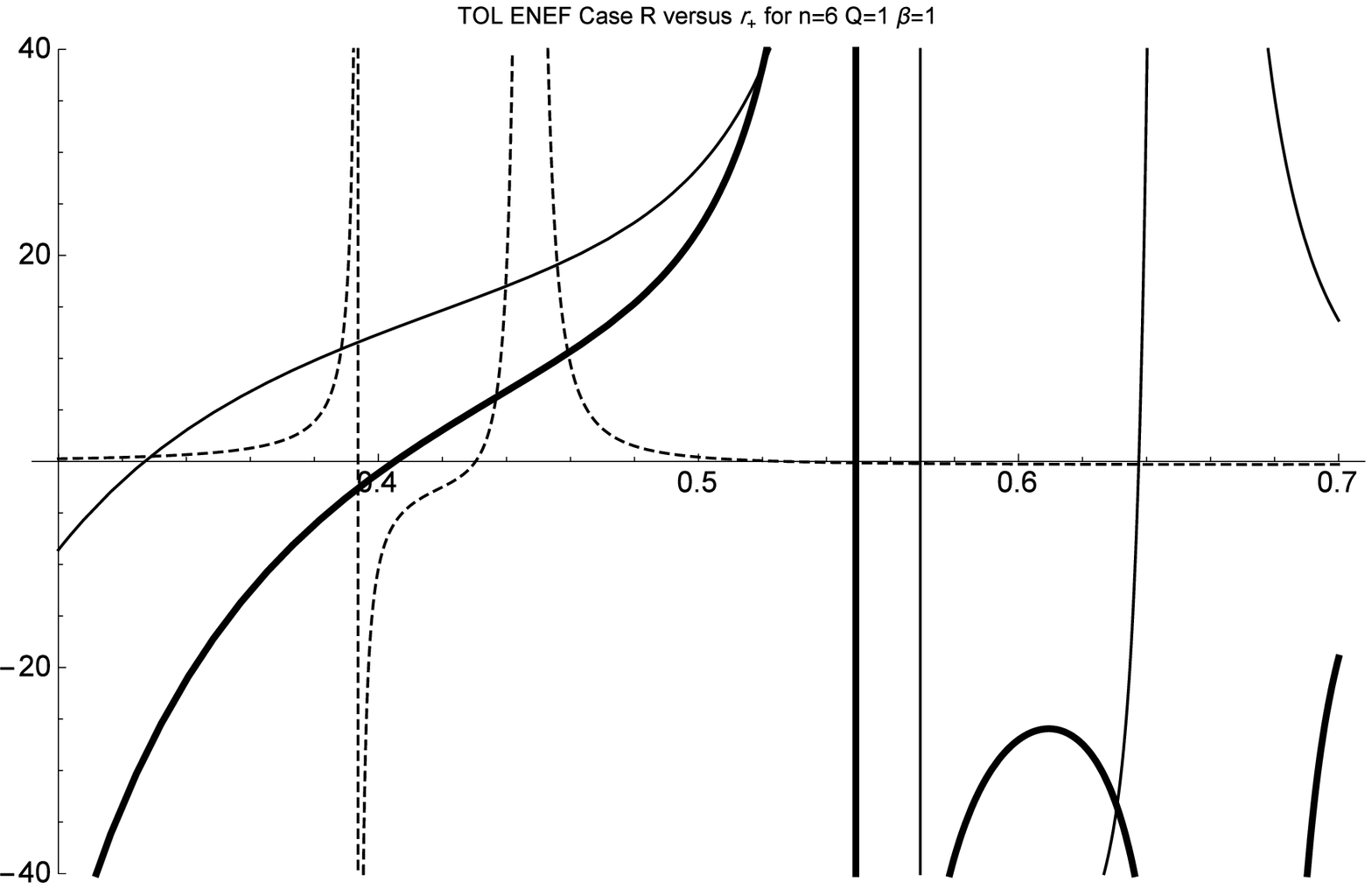} & \epsfxsize=7.5cm %
\epsffile{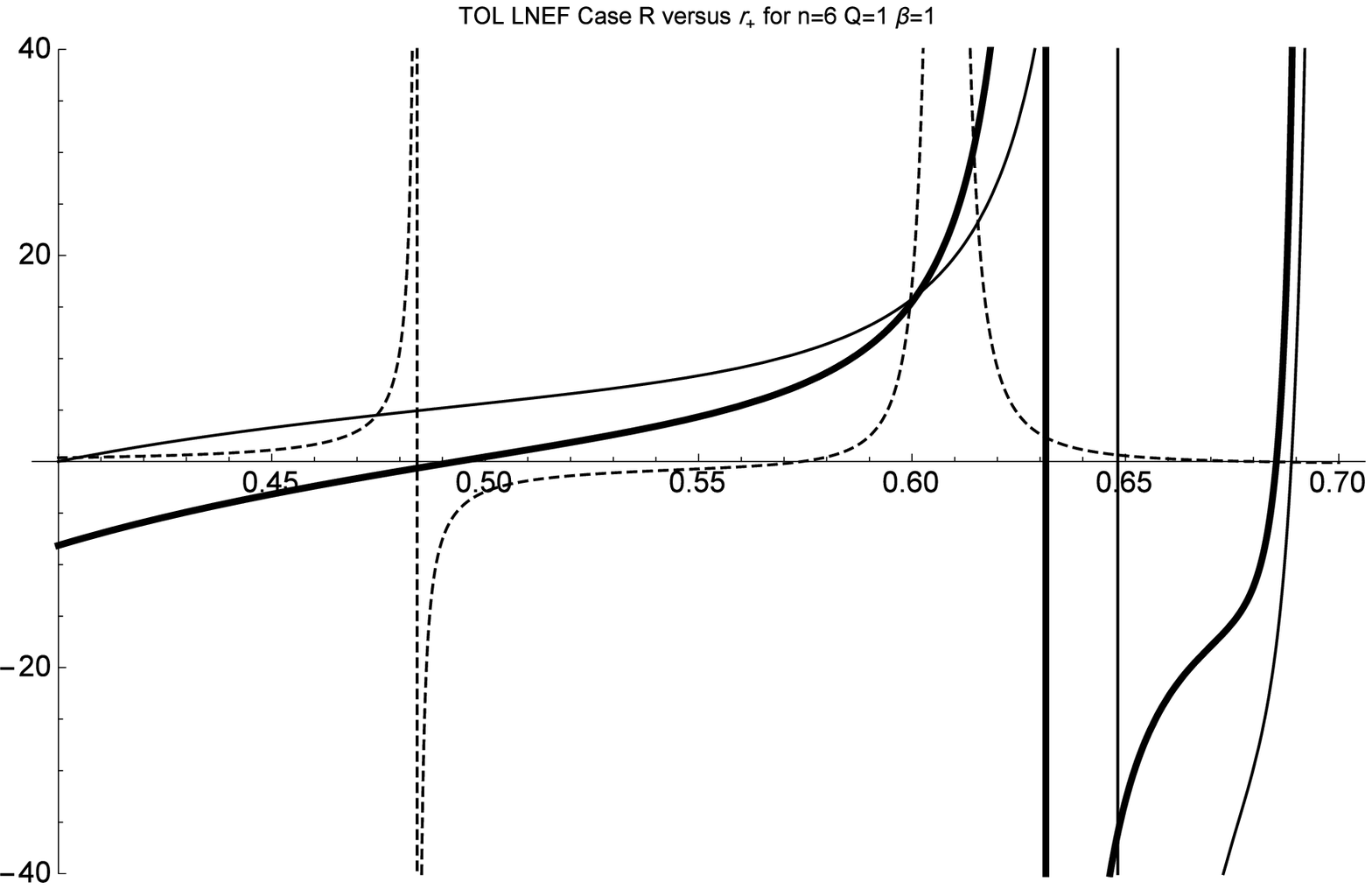}
\end{array}
$%
\caption{\textbf{"TOL case: ENEF (left) and LNEF (right)
branches:"} Heat capacity (up) and Geometric Ricci scalar (down)
versus $r_{+}$ for $n=6$, $\Lambda=-1$, $Q=1$, $\protect\beta=1$
and $\protect \alpha =0.001$ (solid line), $\protect\alpha =0.1$
(bold line) and $\protect\alpha =1$ (dashed line). } \label{HEAT8}
\end{figure}

\begin{figure}[tbp]
$%
\begin{array}{cc}
\epsfxsize=7.5cm \epsffile{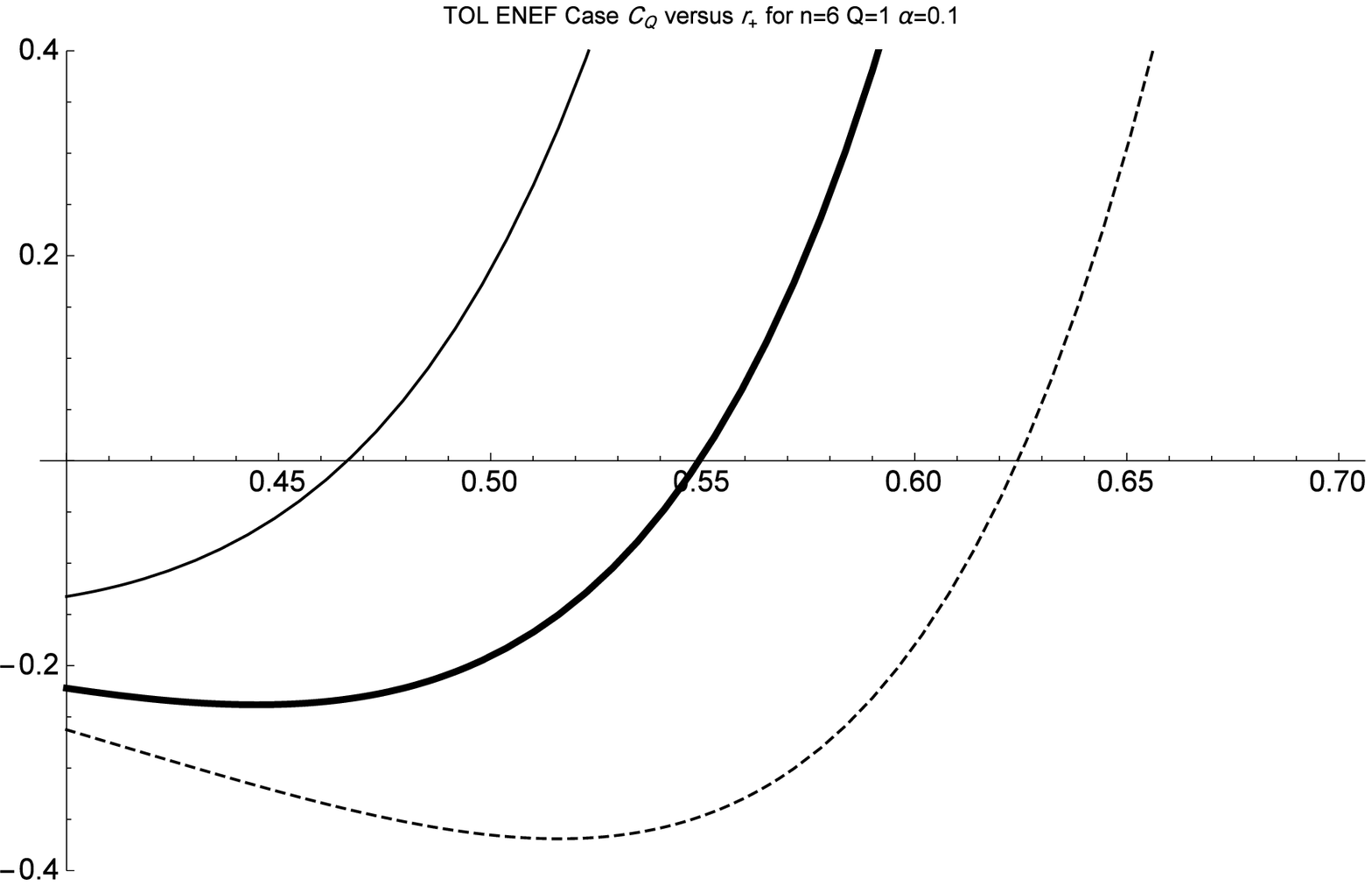} & \epsfxsize=7.5cm %
\epsffile{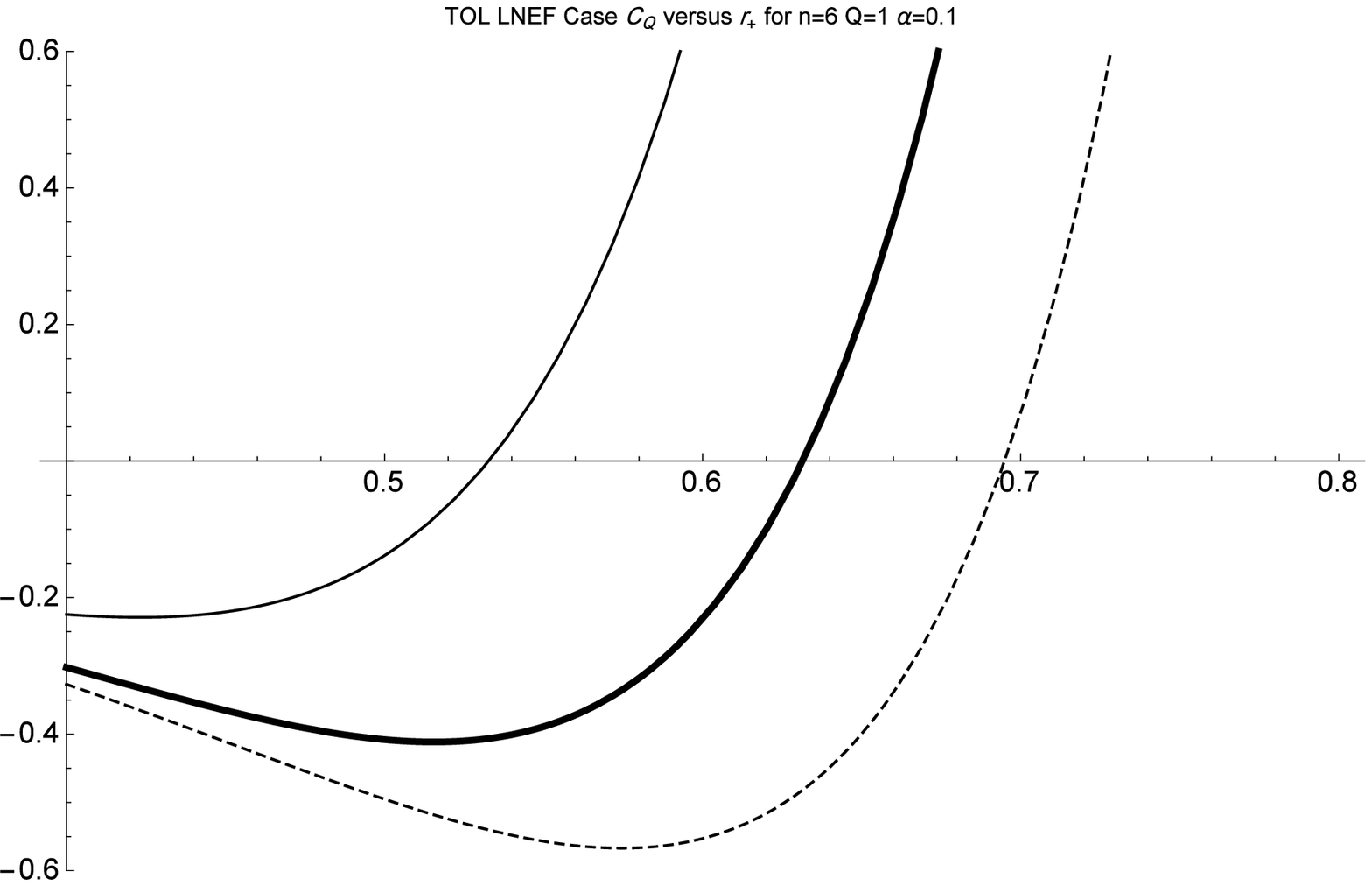}  \\
\epsfxsize=7.5cm \epsffile{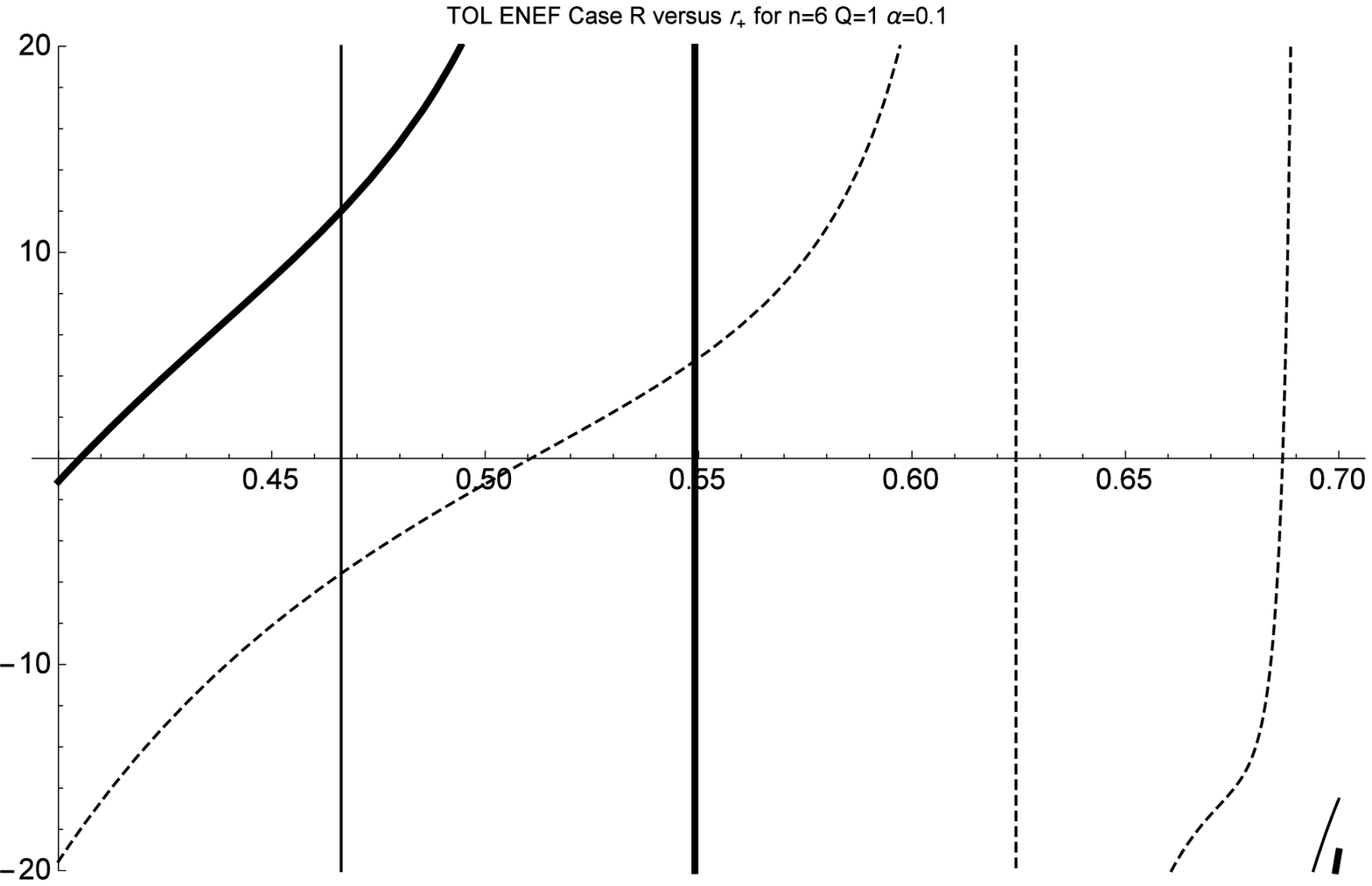} & \epsfxsize=7.5cm %
\epsffile{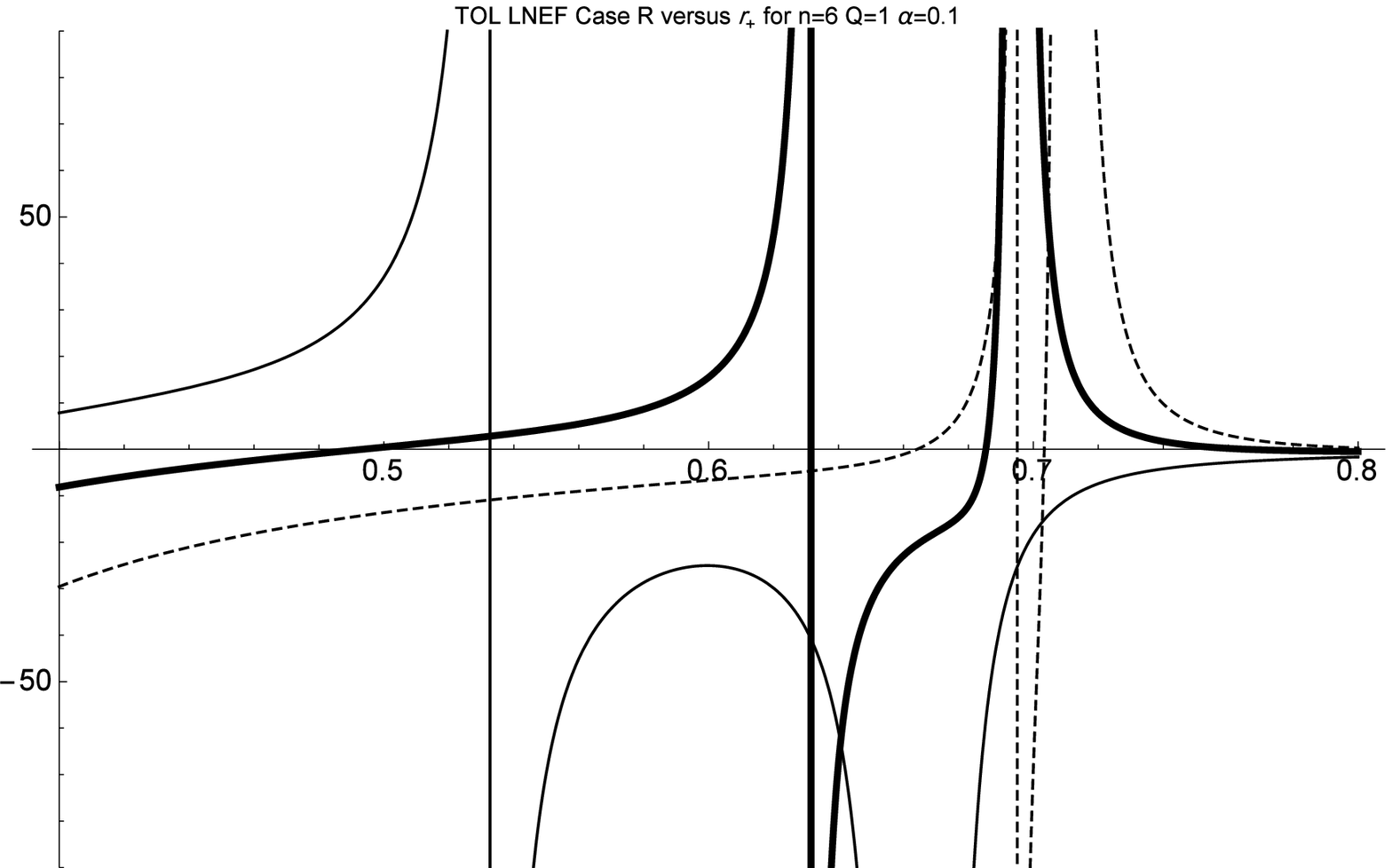}
\end{array}
$%
\caption{\textbf{"TOL case: ENEF (left) and LNEF (right)
branches:"} Heat capacity (up) and Geometric Ricci scalar (down)
versus $r_{+}$ for $n=6$, $\Lambda=-1$, $Q=1$,
$\protect\alpha=0.1$ and $\protect \beta =0.5$ (solid line),
$\protect\beta =1$ (bold line) and $\protect\beta =2$ (dashed
line). } \label{HEAT9}
\end{figure}

Another new method of describing the phase transitions of a
thermodynamical systems is the concept of geometry in
thermodynamics. In this method one may regard the curvature
singularities as the phase transition and so the curvature can be
interpreted as a system interaction. In order to calculate the
heat capacity and the curvature singularity using the GTD method,
we calculate the finite mass as a function of entropy and electric
charge, $M(S,Q)$. One may use Eqs. (\ref{Entropy}), (\ref{TempE})
((\ref{TempGB}) or (\ref{TempTOL}) ) and (\ref{Heat}) to obtain
\begin{equation}
C_{Q}=\frac{T}{\left( \frac{\partial ^{2}M}{\partial S^{2}}\right) _{Q}}=%
\frac{T\left( \frac{\partial S}{\partial r_{+}}\right) _{Q}}{\left( \frac{%
\partial T}{\partial r_{+}}\right) _{Q}}.  \notag
\end{equation}%
Although an analytical expression for $C_{Q}$ is too large and
thus we cannot evaluate its sign analytically, numerical analysis
helps us overcome this problem. In addition, we include Figs.
\ref{HEAT1}-\ref{HEAT9} to provide additional clarification.
Numerical calculations show that there is a lower limit on the
horizon radius for the stable AdS solutions. This means that in
order to have a stable solution, we should set the horizon radius
higher than a lower bound ($r_{+}>r_{min}$) \cite{HendiDehghani}
(we should note that this statement is valid for asymptotically
AdS solutions with spherical horizons).

Now we would like to study the phase transition, which occurs at
$r_{+}=r_{min}$. Up diagrams of Figs. \ref{HEAT1}-\ref{HEAT9} show
that regardless of the value of $\alpha $, $r_{min}$ increases as
the nonlinearity parameter, $\beta$ increases.

Considering the effects of higher orders of Lovelock gravity, we
find that the behavior of Fig. \ref{HEAT4} is different from that
in Figs. \ref{HEAT6} and \ref{HEAT8}. By examining these figures
(and numerical calculations), we find that, regardless of the
values of $q$, $\Lambda$ and $\beta$, the root of the heat
capacity does not change for various choices of $\alpha$ in
$5$-dimensional GB gravity. However, for higher dimensional
solutions of GB gravity, an increase of $\alpha$ leads to a
decrease of $r_{min}$. One can check that for independent Lovelock
coefficients ($\alpha_{2}$ and $\alpha_{3}$), the behavior of
$7$-dimensional TOL gravity is the same as that in $5$-dimensional
GB gravity. Here, we note that this unusual behavior may be
expected for $5$-dimensional GB gravity, $7$-dimensional TOL
gravity, $9$-dimensional fourth order Lovelock (FOL) gravity and
so on. Considering an $(n+1)$-dimensional spacetime, the
contribution of the GB term (TOL term) of Lovelock gravity can be
seen for $n \geq 4$ ($n \geq 6$). It has been shown that the
properties of $5$-dimensional GB solutions are slightly different
from higher dimensional solutions (see \cite{Dehghani1} and the
geometrical mass interpretation). In addition, one can obtain the
same specific behavior for $7$-dimensional TOL gravity
\cite{Dehghani2} (we should choice a constant GB parameter,
$\alpha_{2}$, and vary TOL parameter, $\alpha_{3}$ to see this
unusual behavior), $9$-dimension in FOL gravity (we should choice
constant GB and TOL parameters, $\alpha_{2}$ and $\alpha_{3}$ and
vary FOL parameter, $\alpha_{4}$ to see this unusual behavior) and
so on. In other words, it is expected that the GB parameter in
$5$-dimension, TOL parameter in $7$-dimension, FOL parameter in
$9$-dimension have unusual properties and it may be expected that,
unlike higher dimensional cases, they do not change the location
of the vanishing heat capacity (as we mentioned before, this
property is independent of the nonlinearity parameter $\beta$ and
numerical calculations for $\beta \neq 1$ confirm this point). In
this paper, since we are using a special case, in which
$\alpha_{2}$ is related to $\alpha_{3}$, changing $\alpha_{3}$
leads to a change in $\alpha_{2}$. Therefore, for seven dimensions
($n=6$), we cannot fix $\alpha_{2}$ and vary $\alpha_{3}$ to check
the effect of $\alpha_{3}$, and hence we do not see the unusual
behavior of $7$-dimensional TOL gravity.

In order to study the phase transition using the GTD approach, we
follow the method of Quevedo
\cite{Quevedo1,Quevedo2,Quevedo3,Quevedo4,Quevedo5,Quevedo6,Quevedo7}.
We calculate the Ricci scalar of the Legendre invariant of the
Ruppeiner metric and look for its singularities to compare them
with the zeros of the heat capacity. Although the Quevedo method
is straightforward, analytically calculated results are too large.
So, for the sake of brevity, we do not write the long equations of
the Ricci scalars; instead we use the numerical analysis and some
plots to investigate the Ricci scalar's behavior. We plot $R(S,Q)$
as a function of $r_{+}$ ($r_{+}=r_{+}(S)$) and compare it with
the corresponding heat capacity. Comparing up and down diagrams of
Figs. \ref{HEAT1}-\ref{HEAT9} , we find that the singularities of
the Ricci scalar (down diagrams) take place at those points where
the heat capacity vanishes (up diagrams). Hence, both the GTD
method and the usual thermodynamic approach in the canonical
ensemble are in agreement with each other, which confirms that the
black holes undergo a phase transition.

\section{Conclusions}

In this paper, we considered black hole solutions of the Einstein,
GB and TOL gravities with two classes of BI type NLED models. The
main goal of this paper was to discuss the phase transition using
the GTD method and to compare its consequences with the usual heat
capacity in the canonical ensemble. We obtained the thermodynamic
quantities and adopted the method of Quevedo to obtain the Ricci
scalar of the Legendre invariant of the Ruppeiner metric.

Since the analytical calculations and their corresponding
relations were too large, we performed a numerical analysis.
Numerical calculations showed that the singularities of the Ricci
scalar using the GTD method take place at those points where the
heat capacity vanishes in the canonical ensemble. In other words,
we, interestingly, found that both the GTD method and the usual
thermodynamic stability criterion in the canonical ensemble are in
agreement with each other, which confirms that the black holes
undergo a phase transition.

Moreover, we studied the effects of the nonlinearity parameter,
$\beta$, and the Lovelock coefficient on the location of the
critical points. We found that, regardless of the metric
parameters, the location of the critical points increases as the
nonlinearity parameter, $\beta $ increases. In addition, we showed
that although in general increasing the Lovelock coefficient leads
to a decrease in the location of the critical points of the phase
transition, there are some anomaly cases. We found that this
anomaly takes place for $5$-dimensional GB gravity and may be
generalized to $7$-dimensional TOL gravity, $9$-dimensional FOL
gravity and so on. In other words, the location of the vanishing
heat capacity does not change when changing the GB parameter in
$5$-dimension, when changing the TOL parameter in $7$-dimension,
when changing the FOL parameter in $9$-dimension and so on. It is
interesting to study these anomalies with deep physical insight.

As we mentioned, the asymptotically AdS solutions investigated
here (the third order case) contain only one fundamental constant.
An investigation of the third order case with two independent
constants and a generalization of our results to higher orders of
Lovelock gravity with one (or more) fundamental constant(s) are
interesting subjects for future analysis. Finally, we should note
that it is worthwhile to study the phase transition in an extended
phase space \cite{PVworks}, and this interesting work will appear
in a forthcoming publication \cite{PVcriticality}.

\begin{acknowledgements}
We would like to thank the anonymous referee for valuable
suggestions. We also acknowledge S. Panahiyan for reading the
manuscript. We wish to thank the Shiraz University Research
Council. This work has been supported financially by the Research
Institute for Astronomy \& Astrophysics of Maragha (RIAAM), Iran.
\end{acknowledgements}

\end{document}